\documentclass[aps,prb,a4paper,twocolumn,superscriptaddress,showkeys,longbibliography,floatfix]{revtex4-2}
\usepackage[inline]{enumitem}
\usepackage{dcolumn}
\usepackage{comment} 
\usepackage[dvipsnames]{xcolor}
\usepackage{graphicx}
\usepackage{subfigure}
\usepackage{amsmath}
\usepackage{physics}
\usepackage{subcaption}
\usepackage{bm}
\usepackage{hyperref}
\usepackage{blindtext}
\usepackage{ulem}
\usepackage{comment}
\usepackage{bigints}
\usepackage{placeins}
\hypersetup{colorlinks=true, linkcolor=blue, citecolor=blue, filecolor=magenta, urlcolor=blue }

\newif\ifHighlitedChanges
\def\ifHighlitedChanges{\iftrue}
\ifHighlitedChanges

  \def\STRIKE#1{{\color{red}\sout{#1}}}
\else

  \def\STRIKE#1{\relax}
\fi

\begin{document}
\title{Projective symmetry group classification of fermionic \texorpdfstring{$Z_{2}$}{Z2}  spin liquids on the dipolar-octupolar pyrochlore magnets}

\author{Rohit Deb}
\email{rohitdeb@students.iisertirupati.ac.in}
\affiliation{Department of Physics, Indian Institute of Science Education and Research (IISER)
Tirupati, Tirupati - 517507, Andhra Pradesh, India}
\author{Sambuddha Sanyal}
\email{sambuddha.sanyal@iisertirupati.ac.in}
\affiliation{Department of Physics, Indian Institute of Science Education and Research (IISER)
Tirupati, Tirupati - 517507, Andhra Pradesh, India}

\date{\today}

\begin{abstract}
We construct and classify $Z_2$ quantum spin liquids (QSLs) of fermionic partons on the dipolar-octupolar (DO) pyrochlore lattice. The DO pseudospin shares the space-group and time-reversal algebra of the effective spin-1/2 Kramers doublet, its projective symmetry group (PSG) yields the same 48 symmetric $Z_2$ classes as the generic fermionic classification. We show, the distinct DO symmetry action, encoded in a set of sublattice-independent operators, selects a physically different set of symmetry-allowed mean-field channels, so that the resulting ansatze, spinon spectra, and observables differ from both the effective spin-1/2 case and the bosonic DO classification. Decomposing the anisotropic XYZ exchange model into all symmetric singlet and triplet hopping and pairing channels and minimizing the mean-field energy self-consistently, we obtain phase diagrams in the ($\tilde{J_x}, \tilde{J_z}$) plane in which only a few $Z_2$ states, several supporting gapless or flat-band spinon spectra, are stabilized. For these states we compute the spinon band structures, the low-temperature specific heat, and the dynamical spin structure factor, and we argue that the power-law specific-heat exponent is a diagnostic of the spinon nodal structure. A near-linear-in-temperature specific heat together with broad continua in the structure factor gives a fermionic-parton account of the gapless, fractionalized thermodynamics reported in DO materials such as $Nd_2ScNbO_7$, complementing the bosonic quantum-spin-ice picture of the Ce-based compounds. The work extends the fermionic-parton classification of DO pyrochlore QSLs beyond the $U(1)$ sector, where $Z_2$ topological order is realized equally naturally by fermionic and bosonic spinons.
\end{abstract}
\date{\today}

\maketitle

\tableofcontents

\section{Introduction}

Quantum spin liquids (QSLs) are long-range-entangled ground states of frustrated magnets that evade magnetic order down to zero temperature and host emergent gauge fields and fractionalized excitations~\cite{balents_spin_2010,savary_quantum_2017,zhou_quantum_2017,broholm_quantum_2020}. Three-dimensional QSLs occupy a special position within this landscape: in contrast to two dimensions, a three-dimensional lattice can support a stable deconfined $U(1)$ gauge theory whose transverse fluctuations survive as a gapless, linearly dispersing ``emergent photon,'' so that the low-energy physics realizes an emergent quantum electrodynamics with fractionalized electric (spinon) and magnetic (monopole) charges~\cite{rau_frustrated_2019,gardner_magnetic_2010}. The archetypal host of this physics is the pyrochlore lattice, the network of corner-sharing tetrahedra formed by the magnetic rare-earth ions in $A_2B_2O_7$ oxides, where $A$ is a rare-earth element and $B$ is a transition metal. Its extensive geometric frustration, together with the strongly anisotropic exchange of spin--orbit-entangled $4f$ moments, stabilizes $U(1)$ Coulomb phases obtained by quantum melting of the classical two-in--two-out ice manifold~\cite{huang_quantum_2014,rau_frustrated_2019}, dubbed as quantum spin ices. These states have been studied by a broad range of theoretical methods, including gauge mean-field theory, functional renormalization group, and parton constructions supplemented by variational Monte Carlo~\cite{chern_pseudofermion_2024,li_symmetry_2017,liu_chiral_2026}, while inelastic neutron scattering on candidate rare-earth pyrochlores reveals the broad excitation continua and the persistent absence of magnetic order that are the experimental fingerprints of fractionalization. The pyrochlore magnets thus provide a rare arena in which a controlled three-dimensional emergent-gauge structure can be realized directly with real materials.

A particularly rich subclass is formed by the rare-earth pyrochlores whose crystal-field ground doublet is of the dipolar--octupolar (DO) type~\cite{huang_quantum_2014}. For a DO doublet, two components of the pseudospin ($S^x$ and $S^z$) transform as magnetic dipoles while the third ($S^y$) transforms as a component of a magnetic octupole, and the most general symmetry-allowed nearest-neighbor Hamiltonian reduces, after a uniform rotation about the local $y$ axis, to an anisotropic XYZ model. This structure is special in two respects: it admits octupolar quantum spin ices, in which the ordering degree of freedom carries no dipole moment and is therefore hidden from conventional neutron scattering, alongside the more familiar dipolar ones; and it places the octupolar exchange on an equal footing with the dipolar couplings, so that a single material family spans a wide swath of the anisotropic phase diagram. Experimentally, the DO pyrochlores divide into two groups. The Ce-based compounds Ce$_2$Zr$_2$O$_7$, Ce$_2$Sn$_2$O$_7$, and Ce$_2$Hf$_2$O$_7$ remain disordered to the lowest temperatures and are leading QSL candidates, with thermodynamic, muon, and neutron measurements consistent with a dipolar or octupolar  $U(1)$ quantum spin ice proximate to the all-in-all-out(AIAO) or ice manifold~\cite{gaudet_quantum_2019,gao_experimental_2019,smith_case_2022,sibille_quantum_2020,bhardwaj_sleuthing_2022,yahne_dipolar_2024}. The Nd-based compounds Nd$_2$Zr$_2$O$_7$ and Nd$_2$ScNbO$_7$ instead order into an AIAO antiferromagnet, but the order coexists with dynamical moment fragmentation and with a low-temperature specific heat carrying a near-linear-in-$T$ component incompatible with conventional magnons~\cite{lhotel_fluctuations_2015,xu_anisotropic_2019,scheie_beyond_2021,smith_experimental_2025}. The Nd family thus displays gapless, fractionalized-looking thermodynamics superimposed on a magnetically ordered background.

Because QSLs possess neither a local order parameter nor, generically, a small expansion parameter, their theoretical description is challenging. The sign problem and the growth of entanglement restrict quantum Monte Carlo and tensor-network methods in three dimensions. The available approaches divide broadly by what they compute. Gauge mean-field theory expands about the spin-ice limit and captures the emergent photon perturbatively in the transverse coupling; the pseudofermion (and pseudo-Majorana) functional renormalization group instead computes unbiased spin susceptibilities in the thermodynamic limit and locates the boundaries between ordered and quantum-disordered regimes. For the DO model specifically, a PFRG study has mapped the XYZ phase diagram and identified the $0$-flux and $\pi$-flux quantum spin ices and the competing AIAO orders~\cite{chern_pseudofermion_2024}. These susceptibility-based methods establish where a spin liquid is stabilized, but they do not construct the fractionalized state itself or classify its emergent gauge structure. The parton construction organized by the projective symmetry group (PSG)~\cite{wen_mean-field_1991} fills exactly this gap: it enumerates the symmetry-distinct ways in which spinons can be endowed with a mean-field band structure, providing one of the few systematic routes to construct, distinguish, and compute the observables of candidate three-dimensional fractionalized phases. It is this complementary, state-constructing viewpoint that we adopt here, which also allows a direct comparison against the PFRG phase diagram of the same DO model.

Within the parton framework, two complementary approaches have been applied to the DO pyrochlores. The first uses bosonic (Schwinger-boson) partons and gauge mean-field theory: bosonic spinons condense to produce the proximate magnetic orders, and this route has yielded the dipolar and octupolar $U(1)$ quantum spin ices, their $0$-flux and $\pi$-flux variants, and the associated symmetry-enriched topological orders~\cite{huang_quantum_2014,li_symmetry_2017,yao_pyrochlore_2020}, together with a Schwinger-boson PSG classification of the symmetric $U(1)$ and $Z_2$ DO liquids~\cite{desrochers_competing_2022}. The second uses fermionic (Abrikosov) partons, whose spinons naturally carry gapless spectra---Dirac nodes, nodal lines, and Fermi surfaces---and therefore generate the power-law thermodynamics and broad continua that a gapped-photon spin ice cannot. On the pyrochlore lattice, the fermionic route has produced the complete PSG classification of symmetric $U(1)$ and $Z_2$ spin liquids for generic spin--orbit-coupled moments~\cite{liu_symmetric_2021}, an analysis of the competing orders proximate to these states~\cite{liu_competing_2019}, and, most recently, a classification and variational-Monte-Carlo study of chiral spin liquids~\cite{liu_chiral_2026}; for the DO doublet specifically, the fermionic construction has so far been carried out only in the uniform $U(1)$ sector~\cite{sahu_u1_2024}. Two considerations motivate completing the fermionic side of the DO problem. In three dimensions the bosonic- and fermionic-charge deconfined phases are genuinely distinct, there is no flux attachment relating them, so the fermionic states cannot be recovered from the bosonic classification and must be enumerated in their own right. The $Z_2$ topological order is realized equally naturally by fermionic and bosonic spinons, so the fermionic $Z_2$ DO liquid, and the gapless, fractionalized thermodynamics they can support, are the natural missing entry in the catalogue.

In this paper we (i) carry out the fermionic PSG classification of $Z_2$ DO quantum spin liquids on the pyrochlore lattice; (ii) show that, although the $48$ symmetric $Z_2$ classes coincide with the generic spin--orbit-coupled results~\cite{liu_symmetric_2021} where the gauge structure is fixed by the space group and time reversal, the distinct DO symmetry action selects a physically different set of symmetry-allowed mean-field channels for each class, so that the DO problem yields mean-field solutions distinct from the effective spin-$\tfrac{1}{2}$ pyrochlore; and (iii) construct, for every class, the most general symmetry-allowed mean-field Hamiltonian in the singlet and triplet hopping and pairing channels, and minimize the mean-field energy self-consistently across the anisotropic $(\tilde{J}_x,\tilde{J}_z)$ plane, comparing directly against the $U(1)$ DO states of Ref.~\cite{sahu_u1_2024}. We find that only a small number of $Z_2$ states minimize the energy, several supporting gapless or flat-band spinon spectra, and for these we compute the spinon dispersions, the low-temperature specific heat, which we interpret as a diagnostic of the spinon nodal structure, and the dynamical spin structure factor.

The remainder of the paper is organized as follows. Section~\ref{sec:hamiltonian} introduces the DO spin Hamiltonian and its XYZ form; Section~\ref{sec:psg} sets up the pyrochlore space group, the DO symmetry action, and the fermionic PSG, and presents the $Z_2$ classification; Section~\ref{sec:mft} constructs the mean-field ans\"atze; Section~\ref{sec:observables} presents the spinon band structures, phase diagrams, specific heat, and dynamical structure factor; Section~\ref{sec:experiment} discusses the connection to DO materials; and Section~\ref{sec:conclusion} concludes. Technical details are presented in the Appendices.

\section{Spin Hamiltonian for dipolar-octupolar pyrochlore}\label{sec:hamiltonian}

In this work, we consider the rare-earth pyrochlore magnets with chemical formula $A_2B_2O_7$ \cite{ross_quantum_2011,champion_er_2003,gardner_magnetic_2010,rau_magnitude_2015}, where both ions $A$ and $B$ form pyrochlore lattices. The $B$ site is occupied by a transition metal ion in a $d^1$ or $d^3$ electronic configuration, while the $A$ site forms a trivalent rare-earth ion with a partially filled $4f$ shell. The origin of the magnetic properties of these materials can be attributed to the strong spin-orbit coupling, which has total angular momentum $\mathbf{J} = \mathbf{L} + \mathbf{S}$ of the $A^{3+}$ ions, with the local crystal-field further splitting $J$ into several doublets . At sufficiently low temperatures, the system can be described in terms of an effective pseudospin-$1/2$ degree of freedom associated with the lowest-energy. The symmetry of the pyrochlore site group allows for two types of Kramers doublets and one non-Kramers doublet \cite{huang_quantum_2014}. The two Kramers doublets follow the standard spin commutation relations $[s_i^\mu, s_j^\nu] = i \delta_{ij} \epsilon^{\mu\nu\lambda} s_i^\lambda$ and are odd under time-reversal symmetry. One of these is the $\Gamma_4$ irreducible representation, which behaves exactly like a conventional spin-$1/2$ and is therefore referred to as an effective spin-$1/2$ . The other Kramers doublet arises from a combination of the one-dimensional irreducible representations $\Gamma_5$ and $\Gamma_6$. This is known as the dipolar-octupolar(DO) doublet, the operators $S_i^x$ and $S_i^z$ transform as magnetic dipoles, while $S_i^y$ transforms as a magnetic octupole \cite{huang_quantum_2014}. The DO doublet's pseudospin components do not mix under the $C_3$ rotational symmetry. The non-Kramers doublet transforms according to the $E_g$ irreducible representation, where the magnetic components behave distinctly  under time reversal: $S^z$ is time-reversal odd, whereas the  $S^\pm$ are even. On a local basis, the most general spin Hamiltonian for a DO doublet can be written as \cite{huang_quantum_2014}:

\begin{equation}
\begin{split}
H_l = \sum_{r_\mu,\, r'_\nu} 
& J_{xx} S^{x}_{r_\mu} S^{x}_{r'_\nu} + J_{yy} S^{y}_{r_\mu} S^{y}_{r'_\nu} + J_{zz} S^{z}_{r_\mu} S^{z}_{r'_\nu} \\
& + J_{xz} \left(S^{x}_{r_\mu} S^{z}_{r'_\nu} + S^{z}_{r_\mu} S^{x}_{r'_\nu}
\right).
\end{split}
\end{equation}

Now if we perform a rotation about the local $y$ axis, the new pseudospin operators are defined as
\begin{align}
\tau^x &= S^x \cos\theta - S^z \sin\theta, \nonumber \\
\tau^y &= S^y, \nonumber \\
\tau^z &= S^x \sin\theta + S^z \cos\theta. \nonumber
\end{align}
In this rotated basis, the Hamiltonian takes the form of an XYZ model,
\begin{equation}
H_l = \sum_{\langle r_\mu,\, r'_\nu \rangle} \sum_{a \in \{x,y,z\}} \tilde{J_a} \, \tau^a_{r_\mu} \tau^a_{r'_\nu}.
\label{Heisenberg}
\end{equation}
The rotation angle satisfies $\tan(2\theta) = \frac{2J_{xz}}{J_{zz} - J_{xx}}$, and the effective exchange couplings are
\begin{align}
\tilde{J_x} &= \frac{J_{zz} + J_{xx}}{2} - \frac{1}{2}\sqrt{(J_{zz} - J_{xx})^2 + 4J_{xz}^2}, \nonumber \\
\tilde{J_y} &= J_{yy}, \nonumber \\
\tilde{J_z} &= \frac{J_{zz} + J_{xx}}{2} + \frac{1}{2}\sqrt{(J_{zz} - J_{xx})^2 + 4J_{xz}^2}. \nonumber
\end{align}

In the subsequent section, we will work in the $\tau$ basis for PSG analysis and mean-field calculations, and return to the spin basis for calculating the dynamical structure factor. The details of local to global basis transformation is given in Appendix~\ref{sec:local_global}.

\section{Solving $Z_{2}$ PSG equations on the pyrochlore lattice}\label{sec:psg}

\subsection{Pyrochlore Lattice: space group and time reversal symmetries}

\begin{figure}[!h]
    	\centering
    	\includegraphics[scale=0.15]{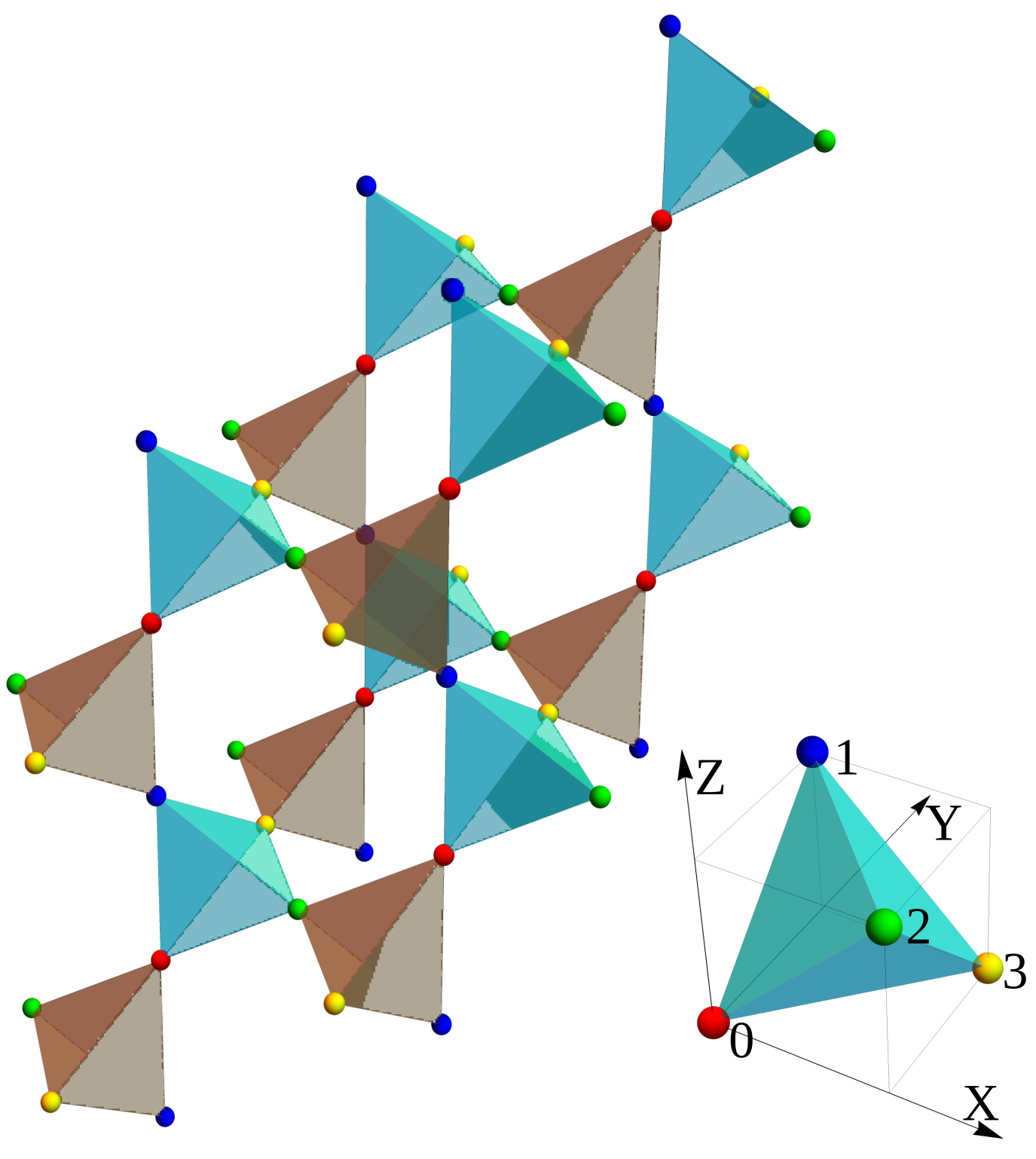}
        \caption{(Top) The pyrochlore lattice. (Inset) The unit cell consists of four sites labeled 0, 1, 2, and 3. The X, Y, and Z represent the global cartesian directions.}
\label{fig:lattice}
\end{figure}

The pyrochlore lattice is characterized as FCC, comprising of four sub-lattices, which are denoted by ${\mu}=0,1,2,3$~(Fig.~\ref{fig:lattice}). The lattice translation vectors of the pyrochlore lattice are given by,
\begin{equation*}
         \textbf{e}_1 = \frac{a}{2}(\hat{y} + \hat{z}) , \textbf{e}_2 = \frac{a}{2}(\hat{x} + \hat{z}), \textbf{e}_3 = \frac{a}{2}(\hat{x} + \hat{y}) .
\end{equation*}
The location of the sublattices are  $\frac{a}{4}(0,0,0),\frac{a}{4}(0,1,1),\frac{a}{4}(1,0,1),\frac{a}{4}(1,1,0)$; where $a$ is the cubic lattice constant. We use the notation $(r_1,r_2,r_3)_\mu$ to refer the location of the sub-lattice $\mu$ in the $(r_1,r_2,r_3)$th unit cell, where
\begin{equation*}
     (r_1,r_2,r_3)_\mu = r_1 \textbf{e}_1 + r_2 \textbf{e}_2 +r_3 \textbf{e}_3 + \frac{1}{2} \textbf{e}_{\mu}.  
\end{equation*}
Here $ \textbf{e}_0=(0,0,0)$.

The following five symmetry operations generate the space group of pyrochlore lattice.
    \begin{equation*}
    \begin{split}
       & T_1 : \mathbf{r_\mu} = (r_1,r_2,r_3)_{\mu} \rightarrow (r_1+1,r_2,r_3)_\mu \\
       & T_2 : \mathbf{r_\mu} = (r_1,r_2,r_3)_{\mu} \rightarrow (r_1,r_2+1,r_3)_\mu \\
       & T_3 : \mathbf{r_\mu}= (r_1,r_2,r_3)_{\mu} \rightarrow (r_1,r_2,r_3+1)_\mu \\
       & \Bar{C_6} : \mathbf{r_\mu}= (r_1,r_2,r_3)_{\mu} \rightarrow \\
       &~~~~~~~~~~~~~~~~~ (-r_3-\delta_{\mu,3},-r_1-\delta_{\mu,1},-r_2-\delta_{\mu,2})_{\Bar{C_6}(\mu)} \\
       & S : \mathbf{r_\mu}= (r_1,r_2,r_3)_{\mu} \rightarrow \\
       &~~~~~ (-r_1-\delta_{\mu,1},-r_2-\delta_{\mu,2},r_1+r_2+r_3+1-\delta_{\mu,0})_{S(\mu)}\\
       & \Bar{C_6}(\mu):
       \begin{pmatrix}
       & 0 &\\
       & 1 &\\
       & 2 &\\
       & 3 &\\
       \end{pmatrix}
       \rightarrow
       \begin{pmatrix}
       & 0 &\\
       & 2 &\\
       & 3 &\\
       & 1 &\\
       \end{pmatrix}, ~{S}(\mu):
       \begin{pmatrix}
       & 0 &\\
       & 1 &\\
       & 2 &\\
       & 3 &\\
       \end{pmatrix}
       \rightarrow
       \begin{pmatrix}
       & 3 &\\
       & 1 &\\
       & 2 &\\
       & 0 &\\
       \end{pmatrix}
    \end{split}
    \end{equation*}
The symmetry transformations are expressed in the pyrochlore sublattice coordinate framework. The operators $T_i$ represent lattice translations along the primitive vectors $\mathbf{e}_i$. In addition, $\bar{C}_6$ denotes a sixfold rotoinversion symmetry about the $[111]$ crystallographic axis. The operator $S$ corresponds to a nonsymmorphic screw symmetry, constructed by combining a twofold rotation about $\mathbf{e}_3$ with a translation of $\frac{1}{2}\mathbf{e}_3$.

The following are the identity relations of the pyrochlore symmetry group.
\begin{equation}
\begin{split}
    T_{i}T_{i+1}T_{i}^{-1}T_{i+1}^{-1}&=1 ,  \forall i
\\
  \Bar{C_6}^6 & = 1\\
  S^2T_3^{-1}& =1\\ \Bar{C_6}T_i\Bar{C_6}^{-1}T_{i+1} & =1 , \forall i\\
   ST_iS^{-1}T_3^{-1}T_i &=1, i=1,2\\
    ST_3S^{-1}T_3^{-1}&=1\\
    (\Bar{C_6}S)^4 &=1\\
     (\Bar{C_6}^3S)^2&=1\\
     \mathcal{T}^2&=-1\\
\mathcal{T}\mathcal{O}\mathcal{T}^{-1}\mathcal{O}^{-1}&=1,~~~~~\mathcal{O}\in {T_1,T_2,T_3,\Bar{C}_6,S}
   \end{split}
   \label{null_conditions}
\end{equation}

where $T$ is translation and $\mathcal{T}$ is time reversal operation.

\subsection{Projective symmetry group }

This section provides a recapitulation of the PSG technique and establishes a systematic framework for classifying all valid QSLs on a given lattice. Liu et al. \cite{liu_symmetric_2021} performed the PSG classification of the pyrochlore lattice for effective dipolar spin $\frac{1}{2}$. As the PSG classification solely depends on the space group and time reversal symmetry of the lattice, it remains the same for the DO spin as well. The underlying key concept lies in the description of the spin degrees of freedom in terms of fermionic spinon/parton degrees of freedom and the constraint of having one particle per site: 

\begin{equation}\label{eq_4}
 \tau^{a}_{\mathbf{r}_\mu} = \frac{1}{2}f^{\dagger}_{\mathbf{r}_\mu \alpha} \sigma^{a}_{\alpha\beta} \, f_{\mathbf{r}_\mu \beta}
\end{equation}

\begin{equation}\label{eq_5}
\sum_{\alpha} f^{\dagger}_{\mathbf{r}_\mu \alpha} f_{\mathbf{r}_\mu \alpha} = 1 \qquad   
\end{equation}

This in turn, provides gauge redundancy to the parton description. The physical effect of $f_{\mathbf{r}_\mu\downarrow}^\dagger$ and $f_{\mathbf{r}_\mu\uparrow}$ is the same, i.e., a one-unit decrease in $\tau_{\mathbf{r}_\mu}^z$. As a result, they can be positioned in a doublet. Similarly, $f_{\mathbf{r}_\mu\downarrow}$ and $-f_{\mathbf{r}_\mu\uparrow}^\dagger$ can produce an equivalent doublet. To manifest this symmetry it is convenient to combine these two doublets and introduce an operator $\Psi_{\mathbf{r}_\mu}=\begin{pmatrix}
f_{\mathbf{r}_\mu\uparrow} & f_{\mathbf{r}_\mu\downarrow}^{\dagger} \\
f_{\mathbf{r}_\mu\downarrow} & -f_{\mathbf{r}_\mu\uparrow}^\dagger\\
\end{pmatrix}$. This allow us to re-write Eq.~\ref{eq_4} and~\ref{eq_5} as 
\begin{equation}
\begin{split}
&\tau_{\mathbf{r}_\mu}^a =\frac{1}{4}\text{Tr}[\Psi^\dagger_{\mathbf{r}_\mu} \sigma^a \Psi_{\mathbf{r}_\mu} ]\\
&\text{Tr}(\Psi_{\mathbf{r}_\mu} \sigma^a \Psi^\dagger_{\mathbf{r}_\mu})=0, ~~a ~\in x,y,z
\end{split}
\label{psi_parton}
\end{equation} 
In this representation, it is easy to see that  spin operators are $SU(2)$ gauge invariant as,
\begin{equation}
    \Psi_{\mathbf{r}_\mu} \rightarrow   \Psi_{\mathbf{r}_\mu} W({\mathbf{r}_\mu}), \tau_{\mathbf{r}_\mu}^a \rightarrow \tau_{\mathbf{r}_\mu}^a, W({\mathbf{r}_\mu}) \in SU(2).
\end{equation}

Since QSLs are fully symmetric states of the magnetic Hamiltonian, we aim to identify the parton mean-field Hamiltonians that preserve all the symmetries of the original spin Hamiltonian. Notably, the symmetry operations acting on the physical spin degrees of freedom are realized projectively in the parton representation. The parton Hamiltonians might be related to one another through gauge transformations which describe the same physical state. Thus, the main objective is to classify all gauge-inequivalent parton Hamiltonians that preserve the full symmetry group of the underlying spin model. The PSG formalism provides a systematic framework to accomplish this classification. Under the space group symmetry operation $O$ and time-reversal operation $ \mathcal{T}$, the spin operator transforms as
\begin{equation}
\begin{split}
    & O : \tau_{\mathbf{r}_\mu}^a \rightarrow U_O \tau_{O(\mathbf{r}_\mu)}^a U_{O}^\dagger \\
    & \mathcal{T} : \tau_{\mathbf{r}_\mu}^a \rightarrow \mathcal{K}U_{\mathcal{T}} \tau_{\mathbf{r}_\mu}^a U_{\mathcal{T}}^{\dagger} \mathcal{K},
\end{split}
\end{equation}
where $\mathcal{K}$ is the complex conjugate operator and $U_O$ for the DO doublets are 
\[
U_{T_i}=1_{2\times2},\quad
U_{\bar{C}_6}=1_{2\times2},\quad
U_S=-i\sigma_y,\quad
U_{\mathcal{T}}=i\sigma_y,
\]
a detailed discussion is given in Appendix~\ref{subsec:do_symmetry}. \\

Due to the local $SU(2)$ gauge redundancy of the parton construction, 
any physical symmetry operation $O$ may be supplemented by a site-dependent 
$SU(2)$ gauge transformation $W({\mathbf{r}_\mu})$. Accordingly, symmetry operations are implemented projectively on the parton as
\begin{equation}
\begin{split}
    \widetilde{O} = G_O O : 
    \quad \Psi_{\mathbf{r}_\mu} 
    &\rightarrow 
    U_O^\dagger 
    \Psi_{O(\mathbf{r}_\mu)} 
    W_{O}(O(\mathbf{r}_\mu)), \\
    \widetilde{\mathcal{T}} = G_{\mathcal{T}} \mathcal{T} : 
    \quad \Psi_{\mathbf{r}_\mu} 
    &\rightarrow 
    U_{\mathcal{T}}^\dagger 
    \Psi_{\mathcal{T}(\mathbf{r}_\mu)} 
    W_{\mathcal{T}}(\mathbf{r}_\mu).
\end{split}
\end{equation}
Here, $\widetilde{O} = G_O.O$ indicates that the projective operation $\widetilde{O}$ consists of the physical symmetry $O$ followed by an associated gauge transformation $G_O$. The collection of all such projective operations $\{\widetilde{O}, \widetilde{\mathcal{T}}\}$ forms the PSG. We determine all PSG solutions in order to classify all symmetry-preserving parton QSL states. These are obtained from the identity relations in eqn. \ref{null_conditions}. Within this framework, such relations are satisfied only up to an additional gauge transformation. Therefore, the null conditions are imposed modulo their corresponding gauge transformations, given by:

\begin{equation}\label{eq_10}
    \widetilde{O}_1 \widetilde{O}_2 \widetilde{O}_3 \cdots 
    = (G_{O_1}. O_1)(G_{O_2}. O_2)(G_{O_3}. O_3)\cdots 
    = \mathcal{G},
\end{equation}

where $\mathcal{G}$ is an element of the invariant gauge group (IGG), the group of all pure gauge transformation that leave the parton Hamiltonian invariant. The local IGG transformation is a subgroup of $SU(2)$ i.e. $Z_2$ or $U(1)$, as it is associated with identity operation it determines the gauge group, thus the classification is often named after it. Using the conjugation relation,

\begin{equation}
     O_i. G_{O_i}. O_i^{-1}\rightarrow G_{O_i}[O_i^{-1}]: \Psi_{\textbf{r}_\mu} \rightarrow \Psi_{\textbf{r}_\mu} W_{O_i}[O_i^{-1}(\textbf{r}_\mu)],
\end{equation}

one can express equation Eq.~\ref{eq_10} purely as a relation between the $SU(2)$ gauge transformation matrices $W({\mathbf{r}_\mu})$, given by
\begin{equation}\label{eq_12}
\begin{aligned}
W_{O_1}[\textbf{r}_\mu]W_{O_2}[O_1^{-1}(\textbf{r}_\mu)] W_{O_3}[O_2^{-1}O_1^{-1}(\textbf{r}_\mu)]...
&= \mathcal{G} \\
&= e^{i \sigma^{3}\chi_{O_1 O_2 O_3...}} \\
&= \eta_{O_1 O_2 O_3...}
\end{aligned}
\end{equation}

The PSG classification is carried out by calculating all algebraic relations of the symmetry group and solving the corresponding $SU(2)$ constraints (provided in Eq.~\ref{eq_12}). Our objective is to identify all gauge-inequivalent solutions, it is necessary to account the relations that are equivalent under a gauge transformation. Accordingly, the group relations must also satisfy
\begin{equation}\label{eq_13}
    (G G_{O_1} O_1 G^{-1})(G G_{O_2} O_2 G^{-1})(G G_{O_3} O_3 G^{-1}) \cdots
    = \mathcal{G},
\end{equation} 
which provides the following transformation relation for the associated $SU(2)$ gauge matrices:
\begin{equation}
    W_{O_i}(\mathbf{r}_\mu)\;\rightarrow\;W(\mathbf{r}_\mu)\,W_{O_i}(\mathbf{r}_\mu)\,
    W^{-1}\!\left(O_i^{-1}(\mathbf{r}_\mu)\right).
\end{equation}

\subsection{Classification result}

We first solved the PSG equations obtained from space-group and time-reversal symmetries and found 48 distinct $Z_{2}$-symmetric spin liquids. The PSG group relations, and hence the number of classes, are fixed by the space group and time reversal, and are therefore identical to the effective spin-$\tfrac{1}{2}$ case of Ref.~\cite{liu_symmetric_2021}. The physical realization of each class, however, is fixed by the DO symmetry action $U_{O}$ obtained in Sec.~\ref{sec:psg}, which differs from the effective spin-$\tfrac{1}{2}$ case and, as shown in Sec.~\ref{sec:mft}, selects a distinct set of symmetry-allowed mean-field channels for each class. The $Z_{2}$ solution has the following form,

\begin{equation}
    \begin{split}
       & W_{T_i}(\mathbf{r}_\mu)= e^{i\sigma^{3}\phi_{T_i}(\mathbf{r}_\mu)},     i=1,2,3 \\
        & W_{\bar{C}_6}(\mathbf{r}_\mu)= g_{\bar{C}_6,\mu}e^{i\sigma^3\phi_{\bar{C}_6}(\mathbf{r}_\mu)} \\
        & W_S(\mathbf{r}_\mu) = g_{S,\mu}e^{i\sigma^3\phi_{S}(\mathbf{r}_\mu)} \\
    \end{split}
\end{equation}
Where
\begin{equation}
    \begin{split}
        & \phi_{T_1}(\mathbf{r}_\mu) = 0, \\
        & \phi_{T_2}(\mathbf{r}_\mu) = -\chi_1r_1, \\
        & \phi_{T_3}(\mathbf{r}_\mu) = \chi_1(r_1-r_2), \\
        & \phi_{\bar{C}_6}(\mathbf{r}_\mu) = -\chi_1r_1(r_2-r_3)+\delta_{\mu,2}\chi_1r_3\\
        &~~~~~~~~~~~~~-r_1[2\chi_{ST_1}+2\chi_1+(\delta_{\mu,2}-\delta_{\mu,3})\chi_1],\\ 
        & \phi_{S}(\mathbf{r}_\mu) = \chi_1[\frac{r_1(r_1+1)}{2}-\frac{r_2(r_2+1)}{2}-r_1r_2]\\
        &~~~~~~~~~~~~~+r_1[\chi_1(\delta_{\mu,1}-\delta_{\mu,2})+(2\chi_1-\chi_{ST_1})]\\
        &+r_2[(2\delta_{\mu,1}-\delta_{\mu,2})\chi_1+3\chi_{ST_1}]+\chi_1r_3[(\delta_{\mu,1}-\delta_{\mu,2})+2]
    \end{split}
\end{equation}

The parameters $\chi_{1}$ and $\chi_{ST_{1}}$ are elements of IGG defined on the right-hand side of the PSG equations obtained from Eq.~\ref{null_conditions}. Moreover, $Z_{2}$ IGG $\eta_{O_1 O_2 O_3...}$, is restricted to only two values, ($\pm1$). Consequently, the allowed values for $\chi_{O_1 O_2 O_3\ldots}$ are limited to $0$ or $\pi$. Given that $W(\mathbf{r}_\mu)$  are $SU(2)$ matrices, the most general form is $W_{O_{i}}(\mathbf{r_{\mu}})=g_{{{O_i},\mu}} e^{i\sigma^{3} \phi_{O_i}(\mathbf{{r_{\mu}}})} $ (where $g_{{{O_i},\mu}}$ and $e^{i\sigma^{3} \phi_{O_i}(\mathbf{r_{\mu}})}$ corresponds to the $SU(2)$ and $Z_{2}$ part of the $W_{O_{i}}(\mathbf{r_{\mu}})$ matrices respectively). The solution of the time-reversal operation is $W_{\mathcal{T}}(\mathbf{r_{\mu}})= i \eta_{\mu}\sigma^{k}$ , where $k$ can be any of the pauli matrices and $\eta_{\mu}= \pm$ is a sublattice dependent sign factor. The complete classification with all the $W(\mathbf{r_{\mu}})$ matrices and time reversal information is given in the Table~\ref{PSG classification}. Different PSG cases described by different values of $\chi_{1}$, $\chi_{ST_{1}}$ , $\chi_{\bar{C_{6}}S}$, $\chi_{S\bar{C_{6}}}$, $\chi_{\bar{C_{6}}}$ , $j$ and $W_{\mathcal{T},0}$ is given in Table~\ref{PSG_cases}.

\begin{table*}[t]

\begin{ruledtabular}
\begin{tabular}{ccccccccccl}
$\chi_1$ & $\chi_{ST_1}$ & $\chi_{\bar{C_{6}}S}$ & $\chi_{S\bar{C_{6}}}$ & $\chi_{\bar{C_{6}}}$ & $j$ &
$\left[ [ \chi_{\mathcal{T} \bar{C_6}} , \chi_{\mathcal{T} S} , W_{\mathcal{T},0} ] \right]^{\ddagger}$ &
$(W_{\bar{C_6},0}, W_{\bar{C_6},1}, W_{\bar{C_6},2}, W_{\bar{C_6},3})$ &
$(W_{S,0}, W_{S,1}, W_{S,2}, W_{S,3})$ \\ \hline

0 or $\pi$ & $\chi_1$ & 0 or $\pi$ & 0 & 0 &
 & $\left[0 ,0, i\sigma^{k}\right]^{\ddagger}$ &
$(1,1,1,1)$ &
$(1,1,e^{i\chi_{\bar{C_6}S}},1)$ \\

0 or $\pi$ & $\chi_1$ & 0 or $\pi$ & $\pi$ & $\pi$ &
 & $\left[\begin{smallmatrix}0,0,i\sigma^{k}\\\pi,0,i\sigma^{k-1}\end{smallmatrix}\right]^{\ddagger}$ &
$(-i\sigma^k,1,1,i\sigma^k)$ &
$(1,1,e^{i\chi_{\bar{C_6}S}},1)$ \\

0 or $\pi$ & $\neg\chi_1$ & 0 & $\pi$ & 0 &
1 & $\left[0 ,\pi, i\sigma^{k-1}\right]^{\ddagger}$ &
$(-e^{i\chi_{\bar{C_6}S}}e^{i(\pi j/3)\sigma^{k-1}},1,1,1)$ &
$(1,i\sigma^k,i\sigma^{k}e^{i(\pi j/3)\sigma^{k-1}},1)$ \\

0 or $\pi$ & $\neg\chi_1$ & 0 & $\pi$ & 0 &
3 & $\left[\begin{smallmatrix}0 ,0 ,i\sigma^{k}\\0,\pi, i\sigma^{k-1}\end{smallmatrix}\right]^{\ddagger}$ &
$(-e^{i\chi_{\bar{C_6}S}}e^{i(\pi j/3)\sigma^{k-1}},1,1,1)$ &
$(1,i\sigma^k,i\sigma^{k}e^{i(\pi j/3)\sigma^{k-1}},1)$ \\

0 or $\pi$ & $\neg\chi_1$ & $\pi$ & $\pi$ & 0 &
0 & $\left[\begin{smallmatrix}0 ,0 ,i\sigma^{k}\\0,\pi, i\sigma^{k-1}\end{smallmatrix}\right]^{\ddagger}$ &
$(-e^{i\chi_{\bar{C_6}S}}e^{i(\pi j/3)\sigma^{k-1}},1,1,1)$ &
$(1,i\sigma^k,i\sigma^{k}e^{i(\pi j/3)\sigma^{k-1}},1)$\\

0 or $\pi$ & $\neg\chi_1$ & $\pi$ & $\pi$ & 0 &
2 & $\left[0,\pi, i\sigma^{k-1}\right]^{\ddagger}$ &
$(-e^{i\chi_{\bar{C_6}S}}e^{i(\pi j/3)\sigma^{k-1}},1,1,1)$ &
$(1,i\sigma^k,i\sigma^{k}e^{i(\pi j/3)\sigma^{k-1}},1)$\\

0 or $\pi$ & $\neg\chi_1$ & 0 or $\pi$ & 0 & $\pi$ &
 & $\left[\begin{smallmatrix}0,0,i \sigma^{k}\\\pi,\pi,i\sigma^{k-1}\end{smallmatrix}\right]^{\ddagger}$ &
$(i\sigma^k,1,1,i\sigma^k)$ &
$(1,i\sigma^{k},e^{i \chi_{\bar{C_6}S}}i\sigma^{k},1)$ \\

0 or $\pi$ & $\neg\chi_1$ & 0 or $\pi$ & $\pi$ & $\pi$ &
0 & $\left[0,\pi,i\sigma^{k-1}\right]^{\ddagger}$ &
$(e^{i(\pi/6)\sigma^{k-1}},1,1,i\sigma^{k-1})$ &
$(1,i\sigma^k,W^{\pi\pi_0}_{S,2},1)$ \\

0 or $\pi$ & $\neg\chi_1$ & 0 or $\pi$ & $\pi$ & $\pi$ &
1 & $\left[\begin{smallmatrix}0,\pi,i \sigma^{k-1}\\ \pi,0,i\sigma^{k} \\ \pi,\pi,i\sigma^{k+1}\end{smallmatrix}\right]^{\ddagger}$ & $(-i\sigma^{k-1},1,1,i\sigma^{k-1})$ &
$(1,i\sigma^k,e^{i \chi_{\bar{C_6}S}}i\sigma^{k},1)$ \\

\end{tabular}
\end{ruledtabular}
\caption{The 48 $Z_{2}$ PSG cases with space group and time-reversal symmetry. Each line in the table corresponds to a different PSG case; $\left[ \right]^{\ddagger}$ represents the corresponding time reversal symmetry. The parameters $\chi_1$, $\chi_{ST_1}$, $\chi_{\bar{C_6}S}$, $\chi_{S\bar{C_6}}$, and $\chi_{\bar{C_6}}$ are discrete elements of the IGG, and $j$ is an extra parameter introduced during solving the PSG equations. The notation $\neg\chi_1$ means $\pi-\chi_{1}$ and all the discrete element of IGG can take either $0$ or $\pi$, as it is a $Z_{2}$ class.The time-reversal parameters $\chi_{\mathcal{T} \bar{C_6}}$ and $\chi_{\mathcal{T} S}$ are related to sublattice ($\eta_\mu$)  by $(\eta_0, \eta_1, \eta_2, \eta_3) = (1, \eta_{\mathcal{T} \bar{C_6}}\eta_{\mathcal{T} S}, \eta_{\mathcal{T} S}, \eta_{\mathcal{T} S})$. The ``k" in time-reversal symmetry can be any of the Pauli matrices. We define the shorthand notation $W^{\pi\pi_0}_{S,2} = e^{i\chi_{\bar{C_6} S}}\, i\sigma^{k+1}\, e^{i(\pi/6)\sigma^{k-1}}$.}
\label{PSG classification}
\end{table*}

\begingroup
\setlength{\tabcolsep}{6 pt}
\renewcommand{\arraystretch}{1.3} 
\begin{center}
\begin{table}
    \begin{tabular}{c c c c c c c c}
       Case No.  & $\chi_1$ & $\chi_{ST_1}$ & $\chi_{\bar{C_{6}}S}$ & $\chi_{S\bar{C_{6}}}$  & $\chi_{\bar{C_{6}}}$ & $j$ & $ W_{\mathcal{T},0}$  \\
       \hline
        1        & 0       &  0            & 0            &  0         & 0   &  & $i\sigma^{k}$   \\
        2        & 0       &  0            & $\pi$            &  0         & 0   &  & $i\sigma^{k}$   \\
        3        & $\pi$       &  $\pi$           & 0            &  0         & 0   &  & $i\sigma^{k}$   \\
        4        & $\pi$       &  $\pi$           & $\pi$            &  0         & 0   &  & $i\sigma^{k}$   \\
        5        & 0       &  0            & 0            &  $\pi$         &  $\pi$  &  & $i\sigma^{k}$   \\
        6        & 0       &  0            & 0            &  $\pi$         &  $\pi$  &  & $i\sigma^{k-1}$   \\
        7        & 0       &  0            & $\pi$            &  $\pi$         &  $\pi$  &  & $i\sigma^{k}$   \\
        8        & 0       &  0            & $\pi$            &  $\pi$         &  $\pi$  &  & $i\sigma^{k-1}$   \\
        9        & $\pi$       &  $\pi$            & 0            &  $\pi$         &  $\pi$  &  & $i\sigma^{k}$   \\
        10        & $\pi$       &  $\pi$            & 0            &  $\pi$         &  $\pi$  &  & $i\sigma^{k-1}$   \\
        11        & $\pi$       &  $\pi$            & $\pi$            &  $\pi$         &  $\pi$  &  & $i\sigma^{k}$   \\
        12        & $\pi$       &  $\pi$            & $\pi$            &  $\pi$         &  $\pi$  &  & $i\sigma^{k-1}$   \\
        13        & 0      &  $\pi$            & 0            &  $\pi$         &  0  & 1 & $i\sigma^{k-1}$   \\
        14        & $\pi$      &  0            & 0            &  $\pi$         &  0  & 1 & $i\sigma^{k-1}$   \\
        15        & 0      &  $\pi$            & 0            &  $\pi$         &  0  & 3 & $i\sigma^{k}$   \\
        16        & 0      &  $\pi$            & 0            &  $\pi$         &  0  & 3 & $i\sigma^{k-1}$   \\
        17        & $\pi$      &  0            & 0            &  $\pi$         &  0  & 3 & $i\sigma^{k}$   \\
        18        & $\pi$      &  0            & 0            &  $\pi$         &  0  & 3 & $i\sigma^{k-1}$   \\
        19        & 0      &  $\pi$            & $\pi$           &  $\pi$         &  0  & 0 & $i\sigma^{k}$   \\
        20        & 0      &  $\pi$            & $\pi$           &  $\pi$         &  0  & 0 & $i\sigma^{k-1}$   \\
        21        & $\pi$      &  0            & $\pi$           &  $\pi$         &  0  & 0 & $i\sigma^{k}$   \\
        22        & $\pi$      &  0            & $\pi$           &  $\pi$         &  0  & 0 & $i\sigma^{k-1}$   \\
        23        & 0      &  $\pi$            & $\pi$           &  $\pi$         &  0  & 2 & $i\sigma^{k-1}$   \\
        24        & $\pi$      &  0            & $\pi$           &  $\pi$         &  0  & 2 & $i\sigma^{k-1}$   \\
        25        & 0      &  $\pi$            & 0           &  0         &  $\pi$  &  & $i\sigma^{k}$   \\
        26        & 0      &  $\pi$            & 0           &  0         &  $\pi$  &  & $i\sigma^{k-1}$   \\
        27        & $\pi$      &  0            & 0           &  0         &  $\pi$  &  & $i\sigma^{k}$   \\
        28        & $\pi$      &  0            & 0           &  0         &  $\pi$  &  & $i\sigma^{k-1}$   \\
        29        & 0      &  $\pi$            & $\pi$           &  0         &  $\pi$  &  & $i\sigma^{k}$   \\
        30        & 0      &  $\pi$            & $\pi$           &  0         &  $\pi$  &  & $i\sigma^{k-1}$   \\
        31        & $\pi$      &  0            & $\pi$           &  0         &  $\pi$  &  & $i\sigma^{k}$   \\
        32        & $\pi$      &  0            & $\pi$           &  0         &  $\pi$  &  & $i\sigma^{k-1}$   \\
        33        & 0      &  $\pi$            & 0           &  $\pi$         &  $\pi$  & 0 & $i\sigma^{k-1}$   \\
        34        & $\pi$      &  0            & 0           &  $\pi$         &  $\pi$  & 0 & $i\sigma^{k-1}$   \\
        35        & 0      &  $\pi$            & $\pi$          &  $\pi$         &  $\pi$  & 0 & $i\sigma^{k-1}$   \\
        36        & $\pi$      &  0            & $\pi$          &  $\pi$         &  $\pi$  & 0 & $i\sigma^{k-1}$   \\
        37        & 0      &  $\pi$            & 0           &  $\pi$         &  $\pi$  & 1 & $i\sigma^{k-1}$   \\
        38        & 0      &  $\pi$            & 0           &  $\pi$         &  $\pi$  & 1 & $i\sigma^{k}$   \\
        39        & 0      &  $\pi$            & 0           &  $\pi$         &  $\pi$  & 1 & $i\sigma^{k+1}$   \\
        40        & $\pi$      &  0           & 0           &  $\pi$         &  $\pi$  & 1 & $i\sigma^{k-1}$   \\
        41        & $\pi$      &  0           & 0           &  $\pi$         &  $\pi$  & 1 & $i\sigma^{k}$   \\
        42        & $\pi$      &  0           & 0           &  $\pi$         &  $\pi$  & 1 & $i\sigma^{k+1}$   \\
        43        & 0      &  $\pi$            & $\pi$           &  $\pi$         &  $\pi$  & 1 & $i\sigma^{k-1}$   \\
        44        & 0      &  $\pi$            & $\pi$           &  $\pi$         &  $\pi$  & 1 & $i\sigma^{k}$   \\
        45        & 0      &  $\pi$            & $\pi$           &  $\pi$         &  $\pi$  & 1 & $i\sigma^{k+1}$   \\
        46        & $\pi$      &  0            & $\pi$           &  $\pi$         &  $\pi$  & 1 & $i\sigma^{k-1}$   \\
        47        & $\pi$      &  0            & $\pi$           &  $\pi$         &  $\pi$  & 1 & $i\sigma^{k}$   \\
        48        & $\pi$      &  0            & $\pi$           &  $\pi$         &  $\pi$  & 1 & $i\sigma^{k+1}$   \\
        
        \hline
    \end{tabular}
    \caption{We identify different PSG cases with different case numbers which are characterized by $\chi$'s $j$ and $W_{\mathcal{T},0}$'s listed in the table. We are using the above convention to refer to different PSG cases.}
    \label{PSG_cases}
\end{table}
\end{center}
\endgroup

\section{Mean-Field Ansatz}\label{sec:mft}

Here we proceed to reformulate the spin Hamiltonian in terms of the various fermionic mean-field channels. After the parton construction $\mathbf{\tau}_i=\frac{1}{2}f_{i\alpha}^{\dagger}\mathbf{\sigma}_{\alpha\beta}f_{i\beta}$, one can rewrite the spin Hamiltonian (shown in Eq.~\ref{Heisenberg}), in terms of a four-fermi term. The goal is to express this quartic fermionic Hamiltonian in terms of all symmetry-allowed quadratic mean-field channels. We begin by describing the spin Hamiltonian, in terms of the fermionic operators 

\begin{equation}\label{eq_17}
    \begin{split}
       & \tilde{J_x} \tau_i^x \tau_j^x = -\frac{\tilde{J_x}}{4}(f_{i\downarrow}^\dagger f_{j\downarrow}^\dagger f_{i\uparrow}f_{j\uparrow}+f_{i\downarrow}^\dagger f_{j\uparrow}^\dagger f_{i\uparrow}f_{j\downarrow}\\
       &+f_{i\uparrow}^\dagger f_{j\downarrow}^\dagger f_{i\downarrow}f_{j\uparrow}+f_{i\uparrow}^\dagger f_{j\uparrow}^\dagger f_{i\downarrow}f_{j\downarrow}) \\
       & \tilde{J_y} \tau_i^y \tau_j^y = \frac{\tilde{J_y}}{4}(f_{i\downarrow}^\dagger f_{j\downarrow}^\dagger f_{i\uparrow}f_{j\uparrow}-f_{i\downarrow}^\dagger f_{j\uparrow}^\dagger f_{i\uparrow}f_{j\downarrow}\\
       &-f_{i\uparrow}^\dagger f_{j\downarrow}^\dagger f_{i\downarrow}f_{j\uparrow}+f_{i\uparrow}^\dagger f_{j\uparrow}^\dagger f_{i\downarrow}f_{j\downarrow}) \\
       & \tilde{J_z} \tau_i^z \tau_j^z = \frac{\tilde{J_z}}{4}(-f_{i\uparrow}^\dagger f_{j\uparrow}^\dagger f_{i\uparrow}f_{j\uparrow}+f_{i\uparrow}^\dagger f_{j\downarrow}^\dagger f_{i\uparrow}f_{j\downarrow}\\
       &+f_{i\downarrow}^\dagger f_{j\uparrow}^\dagger f_{i\downarrow}f_{j\uparrow}-f_{i\downarrow}^\dagger f_{j\downarrow}^\dagger f_{i\downarrow}f_{j\downarrow}). \\
    \end{split}
\end{equation}

Now, recasting Eq.~\ref{eq_17} in terms of all possible quadratic mean-field channels corresponding to the singlet and triplet hopping and pairing terms, [$\hat{\chi}_{ij}=f_{i\alpha}^\dagger \delta_{\alpha\beta}f_{j\beta},\;
\hat{\Delta}_{ij}= f_{i\alpha}[i\sigma^y]_{\alpha\beta}f_{j\beta},\;
\hat{E}_{ij}^{a}=f_{i\alpha}^{\dagger}[\sigma^a]_{\alpha\beta}f_{j\beta},\;
\hat{D}_{ij}^a=f_{i\alpha}[i\sigma^y\sigma^a]_{\alpha\beta}f_{j\beta},\;
\hat{n}_i=f_{i\alpha}^\dagger \delta_{\alpha\beta}f_{i\beta}$ where $\{\alpha,\beta\}=\{\uparrow,\downarrow\}$], and considering the isotropic limit ($\tilde{J_x}=\tilde{J_y}=\tilde{J_z}=\tilde{J}$), one can rewrite the Hamiltonian as,  

\begin{equation}
    \begin{split}
    H=&-\frac{3}{4}\tilde{J}\sum_{\langle i,j \rangle}[\{\hat{\chi}_{ij}^\dagger \hat{\chi}_{ij}+\hat{\Delta}_{ij}^\dagger \hat{\Delta}_{ij} +{\hat{E}_{ij}}^{x\dagger} \hat{E}_{ij}^x+{\hat{E}_{ij}}^{y\dagger} \hat{E}_{ij}^y\\
    &+{\hat{E}_{ij}}^{z\dagger} \hat{E}_{ij}^z+\frac{1}{3}({\hat{D}_{ij}}^{x\dagger} \hat{D}_{ij}^x+{\hat{D}_{ij}}^{y\dagger} \hat{D}_{ij}^y+{\hat{D}_{ij}}^{z\dagger} \hat{D}_{ij}^z)\}\\
    &-\frac{3}{4}\hat{n}_i\hat{n}_j+3\hat{n}_j].\\
    \end{split}
\end{equation}

Upon imposing the condition of half-filling ($f_{i\alpha}^\dagger f_{i\alpha}=1$), the last two terms becomes constants and can therefore be dropped (as their inclusion would only result in a constant shift of the energy spectrum). Then, the mean-field Hamiltonian can be expressed as,
 
\begin{equation}\label{eq_19}
    \begin{split}
    H_{MF}&=-\frac{3}{4}\tilde{J}\sum_{\langle i,j \rangle}[\{{\chi}_{ij}^* \hat{\chi}_{ij}+{\Delta}_{ij}^* \hat{\Delta}_{ij} +{{E}_{ij}^x}^* \hat{E}_{ij}^x+{{E}_{ij}^y}^* \hat{E}_{ij}^y\\
    &+{{E}_{ij}^z}^* \hat{E}_{ij}^z+\frac{1}{3}({{D}_{ij}^x}^* \hat{D}_{ij}^x+{{D}_{ij}^y}^* \hat{D}_{ij}^y+{{D}_{ij}^z}^* \hat{D}_{ij}^z)+h.c.\}\\
    &-\{\abs{\chi_{ij}}^2+\abs{\Delta_{ij}}^2+\abs{E^x_{ij}}^2+\abs{E^y_{ij}}^2+\abs{E^z_{ij}}^2\\
    &+\frac{1}{3}(\abs{D^x_{ij}}^2+\abs{D^y_{ij}}^2+\abs{D^z_{ij}}^2)\}].\\
    \end{split}
\end{equation}    \\

A detailed derivation of Eq.~\ref{eq_19} is given in Appendix~\ref{MF decomposition}.
The most generic mean-field decomposition containing all mean-field channels in the $SU(2)$ basis can be written as,
\begin{equation}
    H_{\text{MF}} = \sum_{i,j} \sum_{\alpha} \text{Tr}\bigl[\sigma^{\alpha}\Psi_{i}u_{ij}^{\alpha}\Psi_{j}^{\dagger}\bigr].
\end{equation}
The index, $\sigma \in {0,x,y,z}$. $\sigma^0$ is the identity matrix and $\sigma^{x/y/z}$ are the Pauli matrices. The $u_{ij}^{\alpha}$ carries the information of mean-field parameters. 
\begin{equation}\label{meanfield-matrix}
    \begin{split}
        & u_{ij}^0 = i \text{Re}(\chi_{ij})\sigma^0 - \text{Re}(\Delta_{ij})\sigma^x - \text{Im}(\Delta_{ij})\sigma^y-\text{Im}(\chi_{ij})\sigma^z \\
        & u_{ij}^x =  \text{Re}(E^x_{ij})\sigma^0 +i\left(\text{Re}(D^x_{ij})\sigma^x + \text{Im}(D^x_{ij})\sigma^y+\text{Im}(E^x_{ij})\sigma^z\right) \\
        & u_{ij}^y =  \text{Re}(E^y_{ij})\sigma^0 +i\left(\text{Re}(D^y_{ij})\sigma^x + \text{Im}(D^y_{ij})\sigma^y+\text{Im}(E^y_{ij})\sigma^z\right) \\
        & u_{ij}^z =  \text{Re}(E^z_{ij})\sigma^0 +i\left(\text{Re}(D^z_{ij})\sigma^x + \text{Im}(D^z_{ij})\sigma^y+\text{Im}(E^z_{ij})\sigma^z\right) \\
    \end{split}
\end{equation}   \\

The matrix $u^{\alpha}_{ij}$ parametrizes the amplitudes of the different mean-field channels. The hermiticity of the mean-field Hamiltonian imposes the condition $u^{\alpha}_{ij} = \left( u^{\alpha}_{ji} \right)^\dagger$. In addition, the PSG and time reversal symmetry further constrains the allowed form of $u^{\alpha}_{ij}$, with the latter acting on it as $u_{ij}^\alpha = -W_{\mathcal{T}}(i) u_{ij}^\alpha W_{\mathcal{T}}^{\dagger}(j)$, where $W_{\mathcal{T}}(i)$ are different for different cases as described in Table~\ref{PSG classification}. A symmetry operation $O$ maps the bond $(i,j)$ to $(i',j')$. Under a PSG transformation $\tilde{O} = G_O . O$, the mean-field Hamiltonian associated with the bond $(i,j)$ transforms to
 
\begin{equation}
\begin{split}
&\text{Tr}[\sigma^\alpha\Psi_{i}u_{ij}^\alpha \Psi_{j}^\dagger] \xrightarrow{\tilde{O}} \text{Tr}[U_{O} \sigma^\alpha U_{{O}}^\dagger \Psi_{i^{\prime}} W_{O}(i^{\prime}) u_{ij}^\alpha W_{O}^{\dagger}(j^{\prime}) \Psi_{j^{\prime}}^\dagger]. \\
\end{split}
\end{equation}

Given that the aim is to construct a PSG symmetric Hamiltonian invariant under the $\tilde{O}$ operation, which can only occur when

\begin{equation}\label{mf-constraint}
\begin{split}
&\text{Tr}[U_{{O}} \sigma^\alpha U_{O}^\dagger \Psi_{i^{\prime}} W_{O}(i^{\prime}) u_{ij}^\alpha W_{O}^{\dagger}(j^{\prime}) \Psi_{j^{\prime}}^\dagger]=\text{Tr}[\sigma^\alpha\Psi_{i^{\prime}}u_{i^{\prime}j^{\prime}}^\alpha\Psi_{j^{\prime}}^\dagger]
\end{split}.
\end{equation}

This relation (Eq.~\ref{mf-constraint}) imposes constraints between the mean-field parameters $u_{ij}^\alpha$ and $u_{i^{\prime}j^{\prime}}^\alpha$. We systematically evaluate the constraints arising from each PSG operation, which reduces the initially large mean-field parameter space to a smaller set of independent parameters. The relation between mean-field parameters on different bond in the unit cell is discussed in Appendix~\ref{Symmetry relation between mean-field parameters}.

\section{Experimentally measurable quantities}\label{sec:observables}

\subsection{Spinon band structure}

To obtain the band structure and dispersion relation for different MF Hamiltonians, we write them in the Bogoliubov-de Gennes (BdG)  basis and calculate the energy in momentum space. The full procedure of energy calculation is described in Appendix~\ref{Enery minimisation}. The energy minimization of all the different PSG cases is summarized in Table~\ref{MFH_summary}. We first focus on the isotropic exchange limit $\tilde{J_x}=\tilde{J_y}=\tilde{J_z}=\tilde{J}$.\\

Here, we plot the energy dispersion of spinons, Bogoliubov quasiparticles, for four different mean-field channels. The band structure corresponding to the cases 16, 19, 37, and 43 is shown in Fig.~\ref{dis_spheat}. In case 16, only the singlet pairing term exists; in case  43, a singlet hopping and pairing term exists; in case 19, only the triplet pairing term exists; and in case 37, both triplet pairing and hopping terms exist. Cases 43 and 37 have degenerate flat bands at the Fermi level. The energy dispersion plots are gapless; we note that this flat bands pinned to the Fermi level require a separate stability analysis.

\begin{table*}[hbt!]
\centering
\tiny
\setlength{\tabcolsep}{11pt}
\renewcommand{\arraystretch}{1.2}
\resizebox{\textwidth}{!}{%
\begin{tabular}{c c c c c}
\hline

Case No. 
& Constraints 
& Non-Zero Parameters 
& $E_{min}$ 
& $C_V \sim T^{\gamma}$ \\

\hline

1 & $\chi=E^a=D^a=0$ & $\Delta$ & $-0.245\tilde{J}$ & $\gamma \sim 0.91$\\
2 & $\Delta=E^a=D^{y}=\chi=0$ & $D^{x},D^{z}$ & $-0.322\tilde{J}$ & $\gamma \sim 0.64$ \\
3(L) & $\chi=E^a=D^a=0$ & $\Delta$ & $-0.244\tilde{J}$ & $\gamma \sim 1.21$ \\
4(L) & $\Delta=E^a=D^{y}=\chi=0$ & $D^{x},D^{z}$ & $-0.140\tilde{J}$ & $\gamma \sim 0.66$ \\
5 & All MF channels zero & - & - & - \\
6 & $\Delta=E^a=D^a=0$ & $\chi$ & $-0.250\tilde{J}$ & $\gamma \sim 1.33$ \\
7 & $\Delta=E^{a}=D^{x}=D^{z}=\chi=0$ & $D^{y}$ & $-0.135\tilde{J}$ & $\gamma \sim 0.80$ \\
8 & $\Delta=E^{x}=E^{y}=D^{x}=D^{z}=0$ & $D^{y},E^{z},\chi$ & $-0.307\tilde{J}$ & $\gamma \sim 0.70$ \\
9(L) & All MF channels zero & - & - & - \\
10(L) & $\Delta=E^a=D^a=0$ & $\chi$ & $-0.256\tilde{J}$ & $\gamma \sim 1.33$ \\
11(L) & $\Delta=E^{a}=D^{x}=D^{z}=\chi=0$ & $D^{y}$ & $-0.128\tilde{J}$ & $\gamma \sim 0.63$ \\
12(L) & $\Delta=E^{x}=E^{y}=D^{x}=D^{z}=0$ & $D^{y},E^{z},\chi$ & $-0.307\tilde{J}$ & $\gamma \sim 0.66$ \\
13 & All MF channels zero & - & - & - \\
14(L) & All MF channels zero & - & - & - \\
15 & $\chi=E^a=D^a=0$ & $\Delta$ & $-0.281\tilde{J}$ & $\gamma \sim 1.01$ \\
16 & $\chi=E^a=D^a=0$ & $\Delta$ & $-0.281\tilde{J}$ & $\gamma \sim 1.01$ \\
17(L) & $\chi=E^a=D^a=0$ & $\Delta$ & $-0.281\tilde{J}$ & $\gamma \sim 1.01$ \\
18(L) & $\chi=E^a=D^a=0$ & $\Delta$ & $-0.281\tilde{J}$ & $\gamma \sim 2.21$ \\
19 & $\chi=E^{a}=D^{y}=\Delta=0$ & $D^{x},D^{z}$ & $-0.190\tilde{J}$ & $\gamma \sim 1.07$ \\
20 & $\chi=E^{x}=E^{z}=D^{y}=\Delta=0$ & $D^{x},D^{z},E^{y}$ & $-0.394\tilde{J}$ & $\gamma \sim 0.97$ \\
21(L) & $\chi=E^{a}=D^{y}=\Delta=0$ & $D^{x},D^{z}$ & $-0.192\tilde{J}$ & $\gamma \sim 1.18$ \\
22(L) & $\chi=E^{x}=E^{z}=D^{y}=\Delta=0$ & $D^{x},D^{z},E^{y}$ & $-0.394\tilde{J}$ & $\gamma \sim 0.89$ \\
23 & All MF channels zero & - & - & - \\
24(L) & All MF channels zero & - & - & - \\
25 & $\chi=E^a=D^a=0$ & $\Delta$ & $-0.182\tilde{J}$ &  $\gamma \sim 3.29$  \\
26 & $E^{a}=D^{a}=0$ & $\chi,\Delta$ & $-0.231\tilde{J}$ & $\gamma \sim 0.97$ \\
27(L) & $\chi=E^a=D^a=0$ & $\Delta$ & $-0.182\tilde{J}$ & $\gamma \sim 3.29$ \\
28(L) & $E^{a}=D^{a}=0$ & $\chi,\Delta$ & $-0.231\tilde{J}$ & $\gamma \sim 0.95$ \\
29 & $\chi=E^{a}=D^{y}=\Delta=0$ & $D^{x},D^{z}$ & $-0.127\tilde{J}$ & $\gamma \sim 0.71$ \\
30 & $\chi=E^{y}=D^{y}=\Delta=0$ & $D^{x},D^{z},E^{x},E^{z}$ & $-0.397\tilde{J}$ & $\gamma \sim 0.86$ \\
31(L) & $\chi=E^{a}=D^{y}=\Delta=0$ & $D^{x},D^{z}$ & $-0.127\tilde{J}$ & $\gamma \sim 0.64$ \\
32(L) & $\chi=E^{y}=D^{y}=\Delta=0$ & $D^{x},D^{z},E^{x},E^{z}$ & $-0.397\tilde{J}$ & $\gamma \sim 1.07$ \\
33 & All MF channels zero & - & - & - \\
34(L) & All MF channels zero & - & - & - \\
35 & All MF channels zero & - & - & - \\
36(L) & All MF channels zero & - & - & - \\
37 & $\chi=\Delta=0$ & $D^{a},E^{a}$ & $-0.365\tilde{J}$ & $\gamma \sim 0.87$ \\
38 & $\chi=\Delta=0$ & $D^{a},E^{a}$ & $-0.401\tilde{J}$ & n.p \\
39 & $\chi=\Delta=D^{x}=D^{z}=E^{x}=E^{z}=0$ & $D^{y},E^{y}$ & $-0.371\tilde{J}$ & $\gamma \sim 1.05$ \\
40(L) & $\chi=\Delta=0$ & $D^{a},E^{a}$ & $-0.365\tilde{J}$ & $\gamma \sim 0.69$ \\
41(L) & $\chi=\Delta=0$ & $D^{a},E^{a}$ & $-0.401\tilde{J}$ & n.p \\
42(L) & $\chi=\Delta=D^{x}=D^{z}=E^{x}=E^{z}=0$ & $D^{y},E^{y}$ & $-0.371\tilde{J}$ & $\gamma \sim 1.04 $  \\
43 & $E^{a}=D^{a}=0$ & $\chi,\Delta$ & $-0.236\tilde{J}$ & $\gamma \sim 1.33$ \\
44 & $E^{a}=D^{a}=0$ & $\chi,\Delta$ & $-0.236\tilde{J}$ & $\gamma \sim 1.33$ \\
45 & All MF channels zero & - & - & - \\
46(L) & $E^{a}=D^{a}=0$ & $\chi,\Delta$ & $-0.251\tilde{J}$ & $\gamma \sim 0.96$ \\
47(L) & $E^{a}=D^{a}=0$ & $\chi,\Delta$ & $-0.251\tilde{J}$ & $\gamma \sim 0.96$ \\
48(L) & All MF channels zero & - & - & - \\

\hline
\end{tabular}
}

\caption{Summary of energy minimization and specific heat information of all mean-field cases (isotropic). The (L) in the bracket specifies larger unit cell. Where n.p stands for no power law behaviour within one temperature decade.}
\label{MFH_summary}

\end{table*}

\subsection{Phase diagram}

In this section, we construct the phase diagrams in the $\tilde{J_x},\tilde{J_z}$ plane (keeping $\tilde{J_y}$ constant) to investigate the energy variations of the MF Hamiltonians in the anisotropic regime. First of all, we choose the MF Hamiltonian to include only the singlet and triplet hopping terms by setting the pairing channel coefficients to zero. In total, we have 20 and 12 different $Z_{2}$ and $U(1)$ \cite{sahu_u1_2024} MF Hamiltonians respectively. Our objective is to determine, for a fixed $\tilde{J_{y}}$, which of the 32 configurations minimizes the energy at each point in the parameter space ($\tilde{J_{x}}, \tilde{J_{y}}, \tilde{J_{z}}$). The phase diagram is shown in Fig.~\ref{phase_diagram}. Next, keeping all the channels active, we perform a similar analysis, and construct a phase diagram. In this case we have 12 $U(1)$ \cite{sahu_u1_2024} and 36 $Z_{2}$ cases to compare, see Fig.~\ref{phase_diagram}. The differing $Z_2$ counts arise from the channel content of the classes: of the $48$ symmetric $Z_2$ classes, $12$ admit no nonzero symmetry-allowed channel at nearest-neighbor order and are inert (entries marked ``All MF channels zero'' in Table~\ref{MFH_summary}); of the remaining $36$, only $20$ support a nonzero hopping channel, which is why the hopping-only comparison involves $20$ $Z_2$ states while the all-channel comparison involves $36$. The structure of the MF Hamiltonian is discussed in Appendix~\ref{MF decomposition}. 
Furthermore, we examined the case where the MF channels are restricted solely to singlet hopping and pairing. Under these conditions, the MF Hamiltonian cannot accommodate anisotropy, because the underlying constraints in the MF decomposition makes it analogous to an isotropic case.

\setcounter{figure}{1}
\begin{figure*}[t]
\centering

\includegraphics[width=0.48\textwidth]{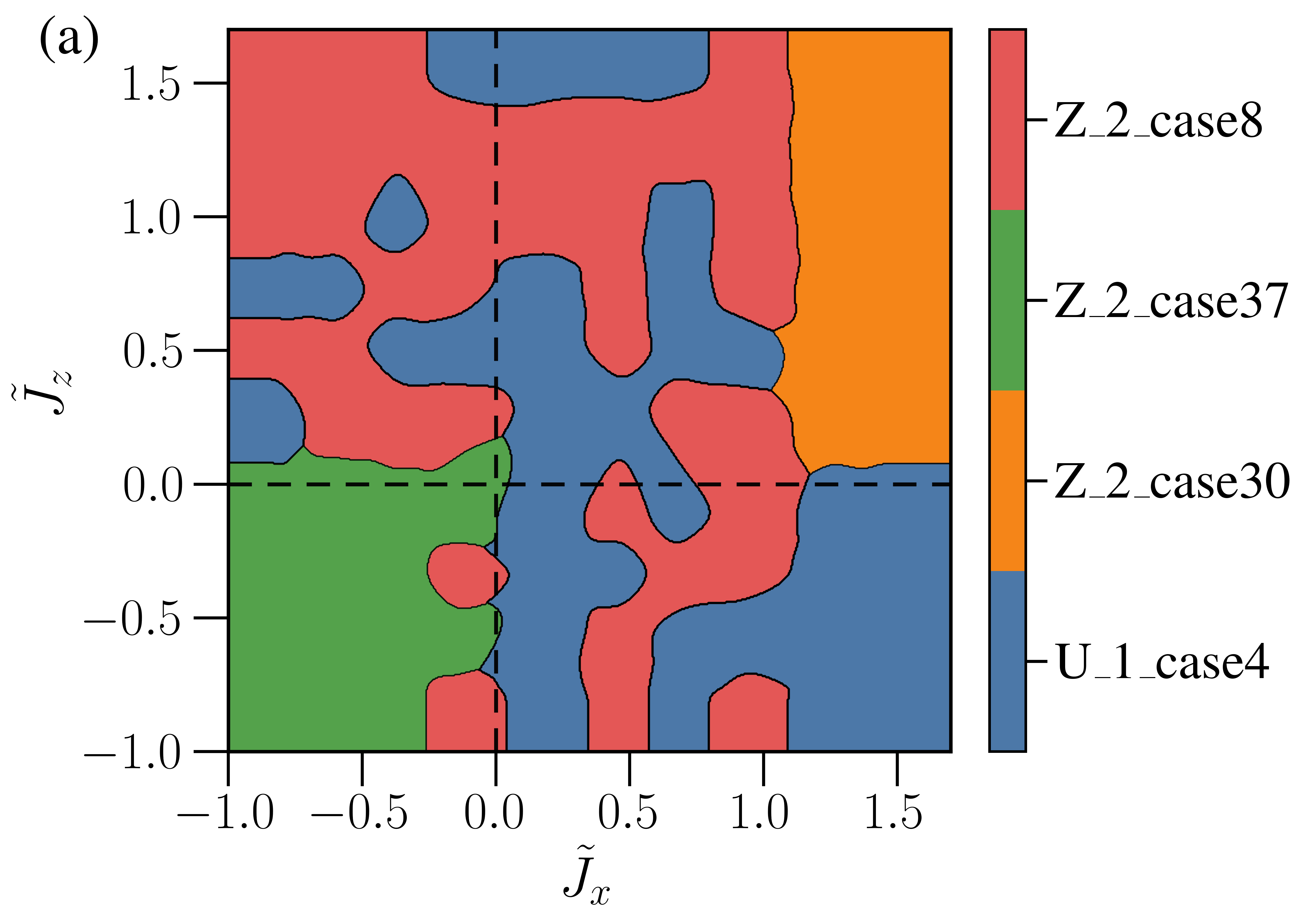}
\hfill
\includegraphics[width=0.48\textwidth]{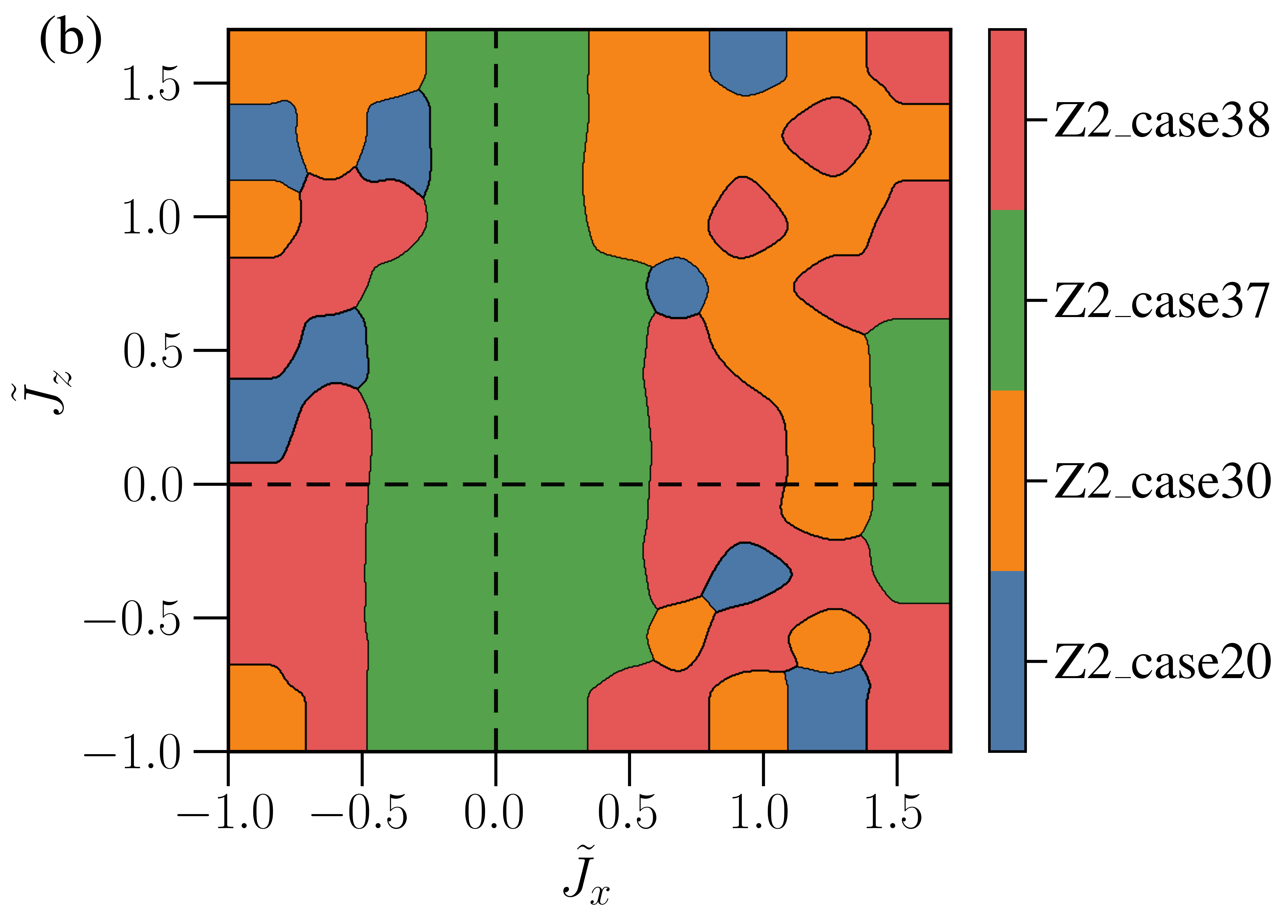}

\caption{
(a)Mean-field phase diagram in the anisotropic-exchange parameter space, restricted to the singlet and triplet hopping channels, with $\tilde{J}_{y}=1$. At each point $(\tilde{J}_{x},\tilde{J}_{y},\tilde{J}_{z})$, the ground-state energies of all 32 mean-field ansatze are compared, and the color denotes the ansatz with the lowest energy. (b) Mean-field phase diagram obtained by including all allowed mean-field channels, again fixing $\tilde{J}_{y}=1$. The ground-state energies of all 48 ansatze are compared at each point in parameter space, with the color indicating the lowest-energy ansatz. The black dashed line denotes the region excluded from the phase diagram, where the exchange couplings cancel the relative weights of the different mean-field channels like $D^x$,$D^z$, $E^x$ and $E^z$.
}
\label{phase_diagram}
\end{figure*}

\subsection{Specific heat}

Here, we discuss the temperature dependence of the specific heat of all PSG cases. The same is governed by low-energy Bogoliubov quasiparticle excitations. The specific heat capacity is given by : 

\[
C_{V} \;=\; \frac{\partial \langle H_{MF} \rangle}{\partial T}
\;=\;
g_{\alpha}(\mathbf{k})\sum_{\mathbf{k}}
\frac{\partial}{\partial T}
\left[
\omega_{\alpha}(\mathbf{k})\, n(\omega_{\alpha}(\mathbf{k})
\right]
\]
\[
=\;
\sum_{\mathbf{k} \in \mathrm{BZ},\, \omega_{\alpha}(\mathbf{k}) > 0}
g_{\alpha}(\mathbf{k}) \,
\frac{\bigl(\omega_{\alpha}(\mathbf{k}) - \mu)^2}
{k_B T^2}
\,
\frac{
\exp\!\left(\dfrac{\omega_{\alpha}(\mathbf{k}) - \mu}{k_B T} \right)
}{
\left[
1 + \exp\!\left(\dfrac{\omega_{\alpha}(\mathbf{k}) - \mu}{k_B T}\right)
\right]^2
}
\]

where $n(\omega_{\alpha})$ is the Fermi-Dirac distribution function and $g_{\alpha}(\mathbf{k})$ is the degeneracy index of the spinon bands. In Fig~\ref{dis_spheat}, the temperature dependence of specific heat is plotted for PSG cases 16, 19, 37 and 43. Cases 19 and 16 exhibit a nearly linear temperature dependence at very low temperatures. The temperature dependence of all other PSG cases is summarized in Table~\ref{MFH_summary}.

\setcounter{figure}{2}

\begin{figure*}[t]
\centering

\includegraphics[width=1\textwidth]{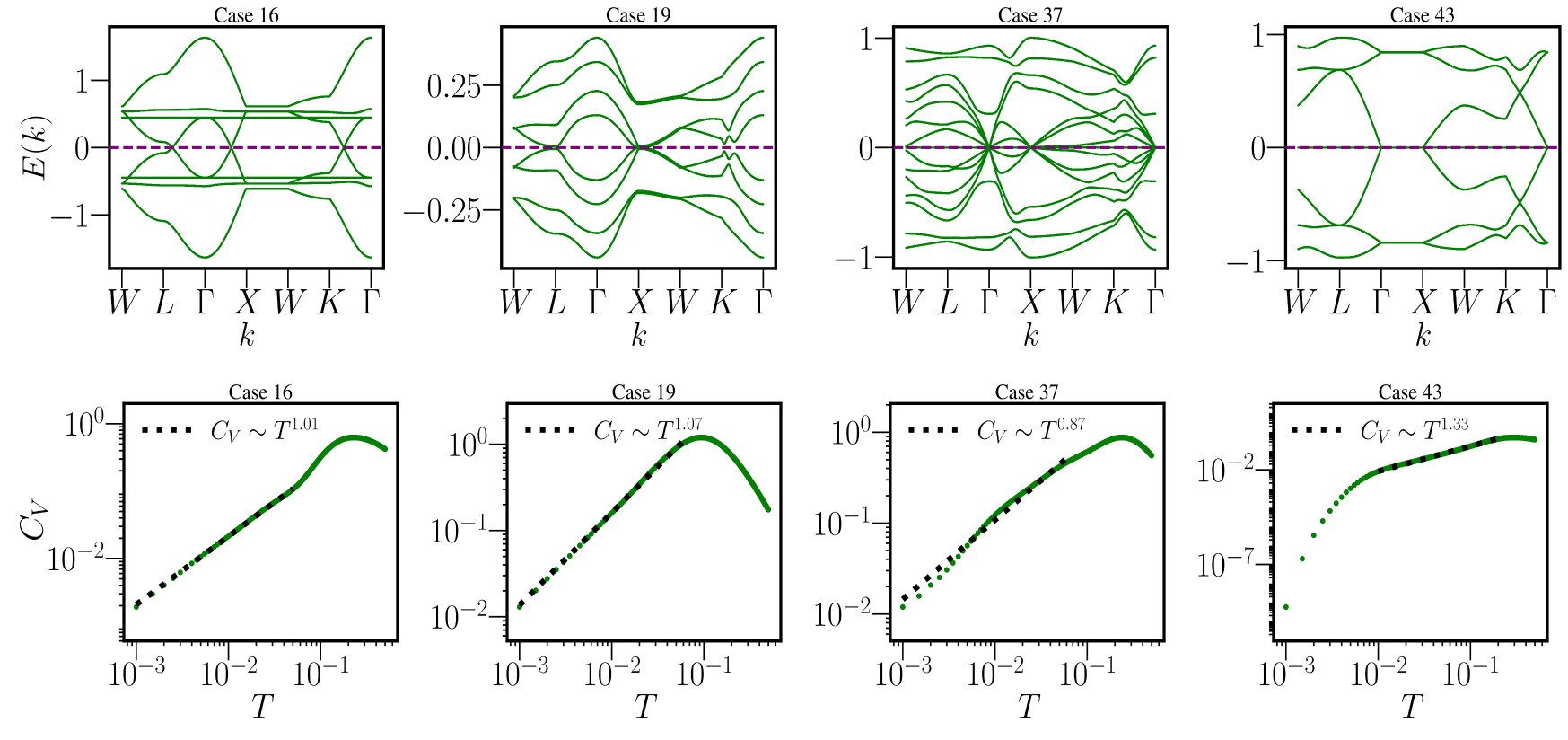}

\caption{Spinon band structures (top) and temperature dependence of the specific heat, $C_V$ (bottom), for different PSG cases 16, 19, 37, and 43 in the isotropic Hamiltonian($\tilde{J_x}=\tilde{J_y}=\tilde{J_z}=\tilde{J}$). The purple dashed line in the band-structure plots denotes the Fermi energy at half-filling. The black dashed lines in the specific heat plots show the corresponding linear fits used to determine the power-law exponent.}
\label{dis_spheat}
\end{figure*}

\begin{figure*}[htbp!]
\centering
\includegraphics[width=\textwidth]{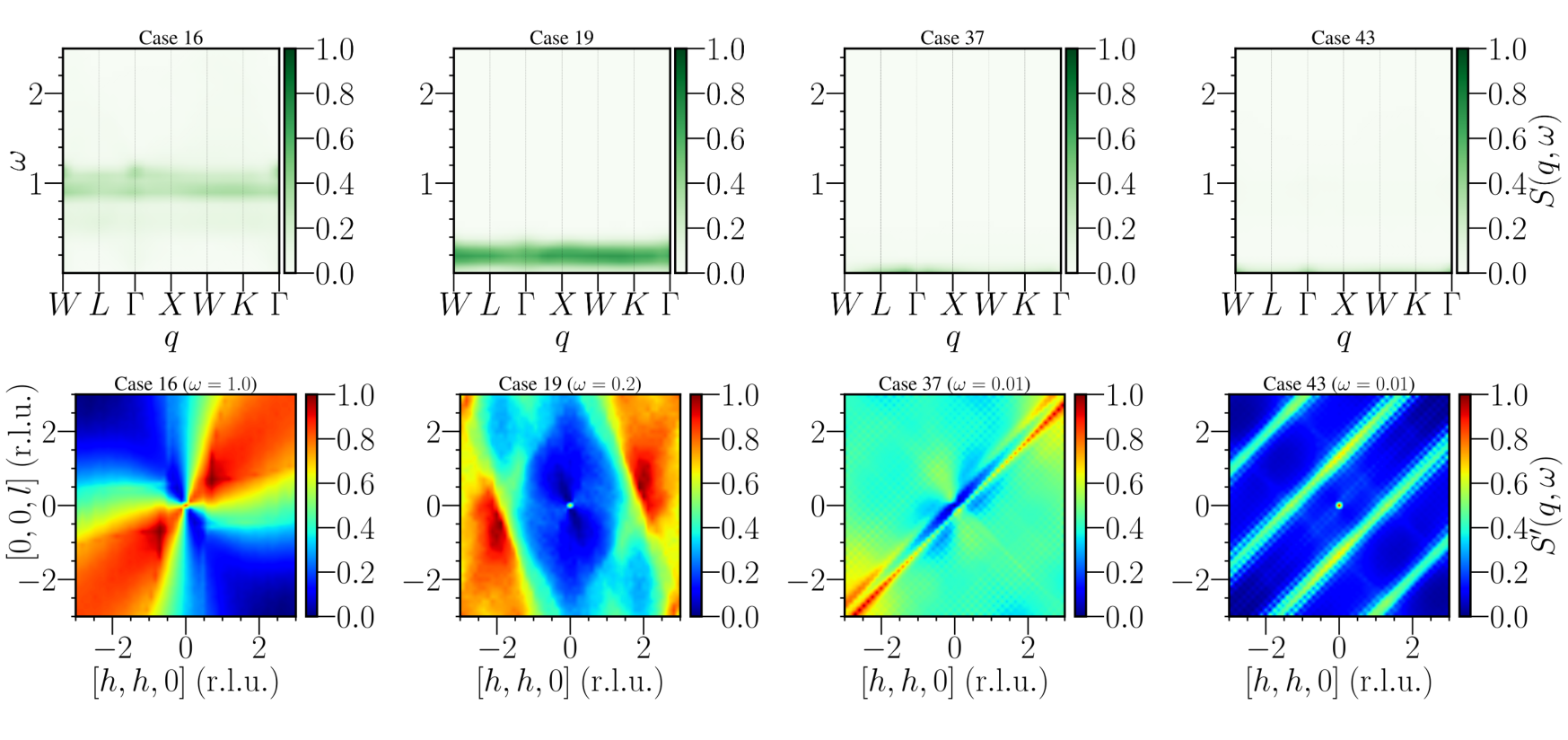}

\caption{Dynamical (top) and static (bottom) structure factors, shown in color scale, along high-symmetry momentum paths and within the high-symmetry $[h,h,l]$ plane for PSG cases 16, 19, 37, and 43 in the isotropic Hamiltonian ($\tilde{J_x}=\tilde{J_y}=\tilde{J_z}=\tilde{J}$).}
\label{st_factor}
\end{figure*}

\subsection{Dynamical spin structure factor}

This section addresses the dynamical spin structure factor, which is analogous to the neutron scattering experiment. It is defined as 

\begin{align}
S_{\mathrm{INS}}(\mathbf{q}, \omega)
&=
\sum_{a,b}
\int_{-\infty}^{\infty} dt \, e^{i \omega t}
\left(
\delta_{ab}
-
\frac{q_a q_b}{|\mathbf{q}|^2}
\right)  \nonumber \\
&\qquad \times
\left\langle
m_a(\mathbf{q}, t)\, m_b(-\mathbf{q}, 0)
\right\rangle .
\end{align}

where the subscript ${\mathrm{INS}}$ means inelastic neutron scattering and the indices $a,b$ run over three different component directions x, y and z. The $m_a(\mathbf{q}, t)$ is defined as

\begin{align}
m_a(\mathbf{q}, t)
=
\frac{1}{\sqrt{N_{\text{u.c.}}}}
\sum_{r_\mu}
e^{i \mathbf{q}\cdot \mathbf{r}_\mu}
\sum_{b}
g^{ab}_{\mu}\,
\tilde{S}^{\,b}_{\mathbf{r}_\mu} \nonumber \\
=
\frac{1}{\sqrt{N_{\text{u.c.}}}}
\sum_{r_\mu}
e^{i \mathbf{q}\cdot \mathbf{r}_\mu}\,
\hat{z}^{\,a}_{\mu}
\sum_{b}
g'_{b}\,
S^{\,b}_{\mathbf{r}_\mu} .
\end{align}

Here, $\tilde{S}^{\,b}_{\mathbf{r}_\mu}$ denotes the spin operator at position
$\mathbf{r}_\mu$ expressed in the global coordinate system.
The matrix $g^{\mu}_{ab}$ is the coupling matrix describing how the
neutron magnetic moment couples to the global spin operator.
The position vector is given by $\mathbf{r}_\mu = \mathbf{r} + \mathbf{d}_\mu$,
where $\mathbf{r}$ is the translation vector of the pyrochlore lattice
and $\mathbf{d}_\mu$ is the corresponding sublattice vector.
The vector $\hat{z}^{\,a}_{\mu}$ defines the local $z$ axis on sublattice $\mu$,
and $S^{\,a}_{\mathbf{r}_\mu}$ represents the spin operator in the local
coordinate system.
The coupling strength of the neutron magnetic moment to the local
spin components $S^{x}$, $S^{y}$ and $S^{z}$ is denoted by
$g' = (g_x, g_y, g_z)$, where $g_x$, $g_y$ and $g_z$ are the respective
coupling constants. \\

Furthermore, we have calculated the structure factor in the experimentally accessible [h,h,l] lattice plane. We examine $S^{\prime}(q,\omega)$ for a particular energy $\omega$ and for each $\omega$, we integrate $S'$ over a small energy window $(\omega -\Delta\omega, \omega +\Delta\omega)$ centered at $\omega$. Here $S'(q, \omega) = \int_{\omega - \Delta \omega}^{\omega + \Delta \omega} d\omega' S(\mathbf{q}, \omega') $.                              \\

The dynamical structure factors and the structure factor calculated along high-symmetry planes in reciprocal space for PSG cases 16, 19, 37, and 43 are presented in Fig.~\ref{st_factor}. The cases 37 and 43 exhibit low-energy excitations arising from the presence of degenerate flat bands at the Fermi level. Moreover, cases 16 and 19 feature a continuous scattering spectrum, which serves as a definitive hallmark of existence of spinon quasiparticles. \\

\begin{figure*}[htbp!] 
    \centering
    
    \includegraphics[width=\textwidth]{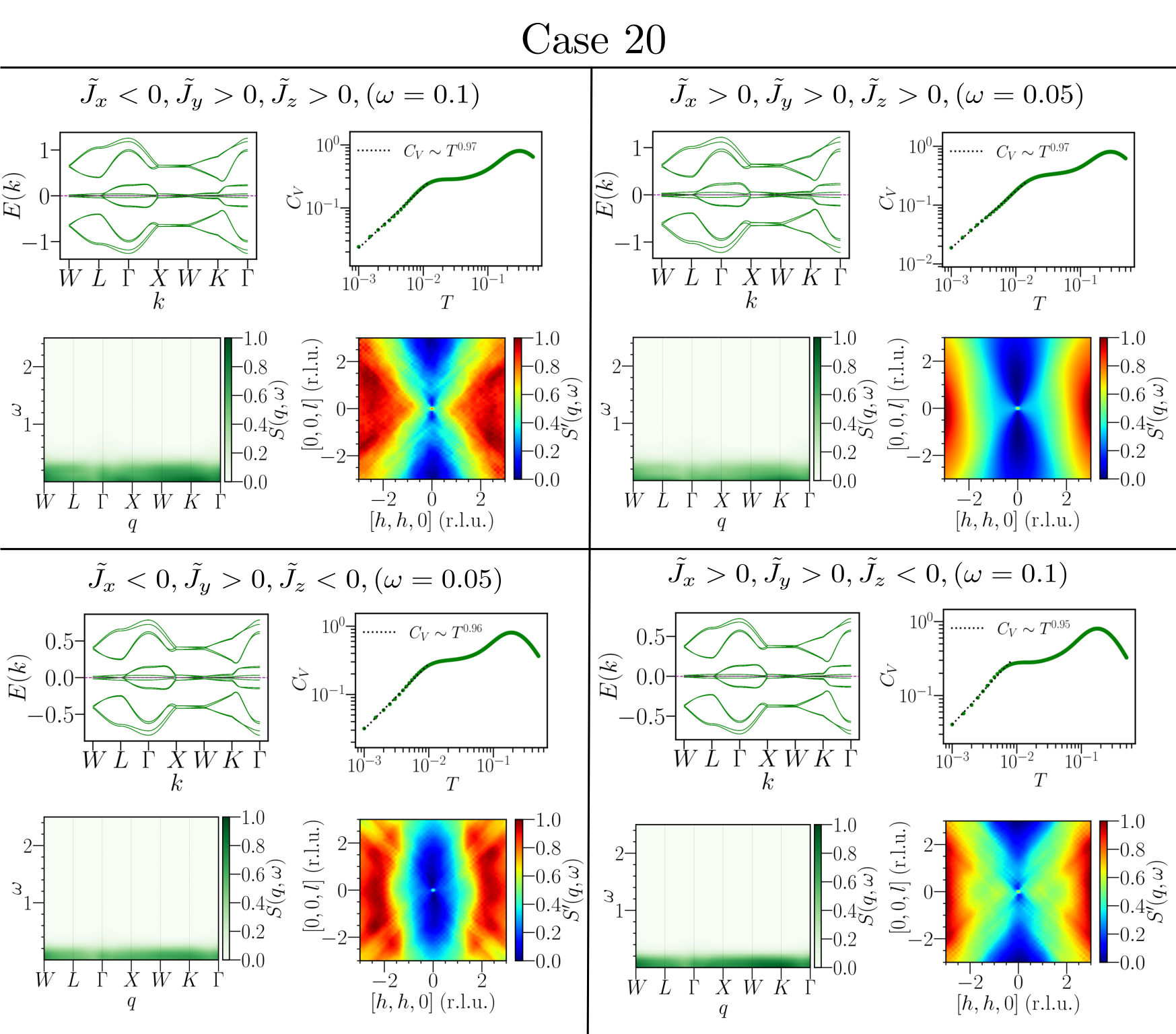}

    \caption{PSG case 20. A flat band at the Fermi energy coexists with other dispersive bands across all four exchange coupling regimes. The low energy $C_V(T)$ is nearly linear ($\gamma \sim 0.95-0.97$) and shows a plateau before its maximum. The $S(\mathbf{q},\omega)$ weight is confined to $\omega \leq 0.3 \tilde{J}$. The $S'(\mathbf{q},\omega)$ cut shows a prominent low-intensity channel along $[0,0,l]$ independent of the exchange coupling regime and a bow-tie-like feature at $\omega = 0.1$ in regimes II and IV.}
    \label{case20}
\end{figure*}

\begin{figure*}[htbp!] 
    \centering
    
    \includegraphics[width=\textwidth]{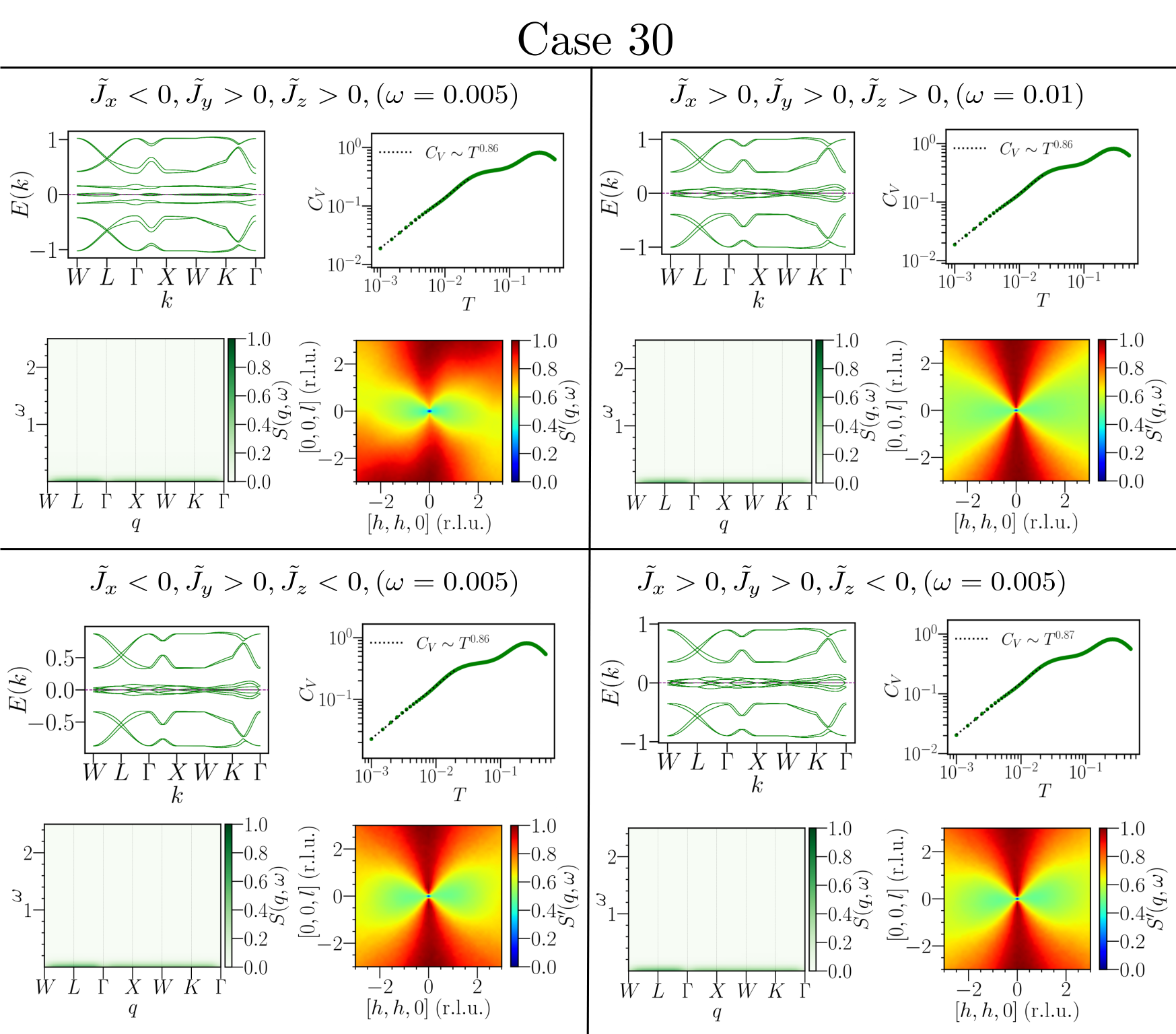}

    \caption{PSG case 30. A flat band at the Fermi energy coexists with other dispersive bands across all four exchange coupling regimes. The low energy $C_V(T)$ is sub-linear ($\gamma \sim 0.86-0.87$) and shows a plateau before its maximum. The $S(\mathbf{q},\omega)$ weight is mostly confined to $\omega \sim 0.005 \tilde{J}$. The $S'(\mathbf{q},\omega)$ cut shows a bow-tie-like feature brightest along $[0,0,l]$ and nearly identical in all four regimes.}
    \label{case30}
\end{figure*}

\begin{figure*}[htbp!] 
    \centering
    
    \includegraphics[width=\textwidth]{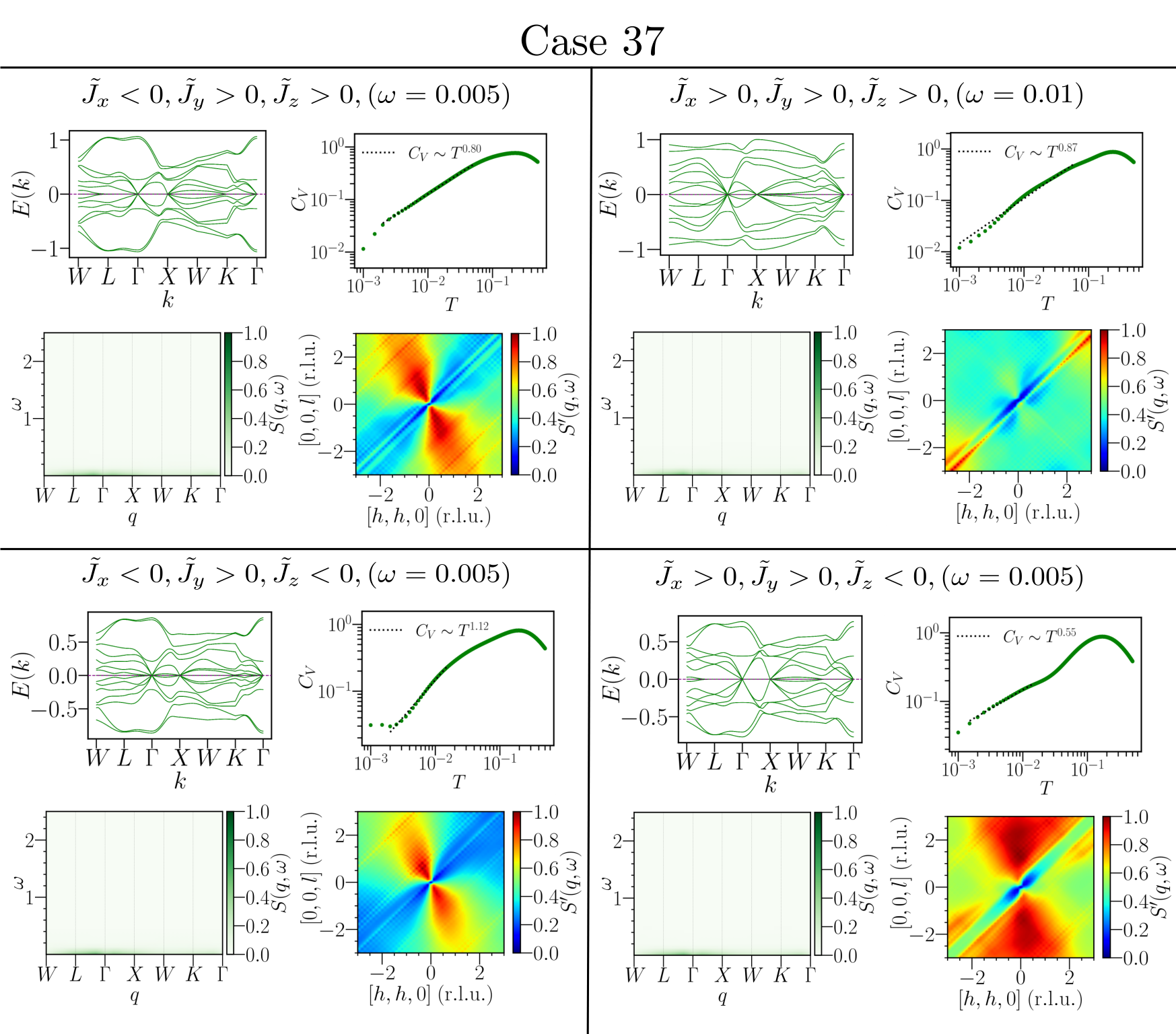}

    \caption{PSG case 37. A flat band at the Fermi energy coexists with other dense, dispersive bands that cross zero energy at several points. The low energy $C_V(T)$ exponent is strongly regime dependent with $\gamma $= 0.87, 0.80, 1.12, and 0.55 for regimes I-IV. The $S(\mathbf{q},\omega)$ weight is mostly confined to $\omega \sim 0.005 \tilde{J}$. The $S'(\mathbf{q},\omega)$ cut shows a pinch-point-like structure in regimes II and III.}
    \label{case37}
\end{figure*}

\begin{figure*}[htbp!] 
    \centering

    \includegraphics[width=\textwidth]{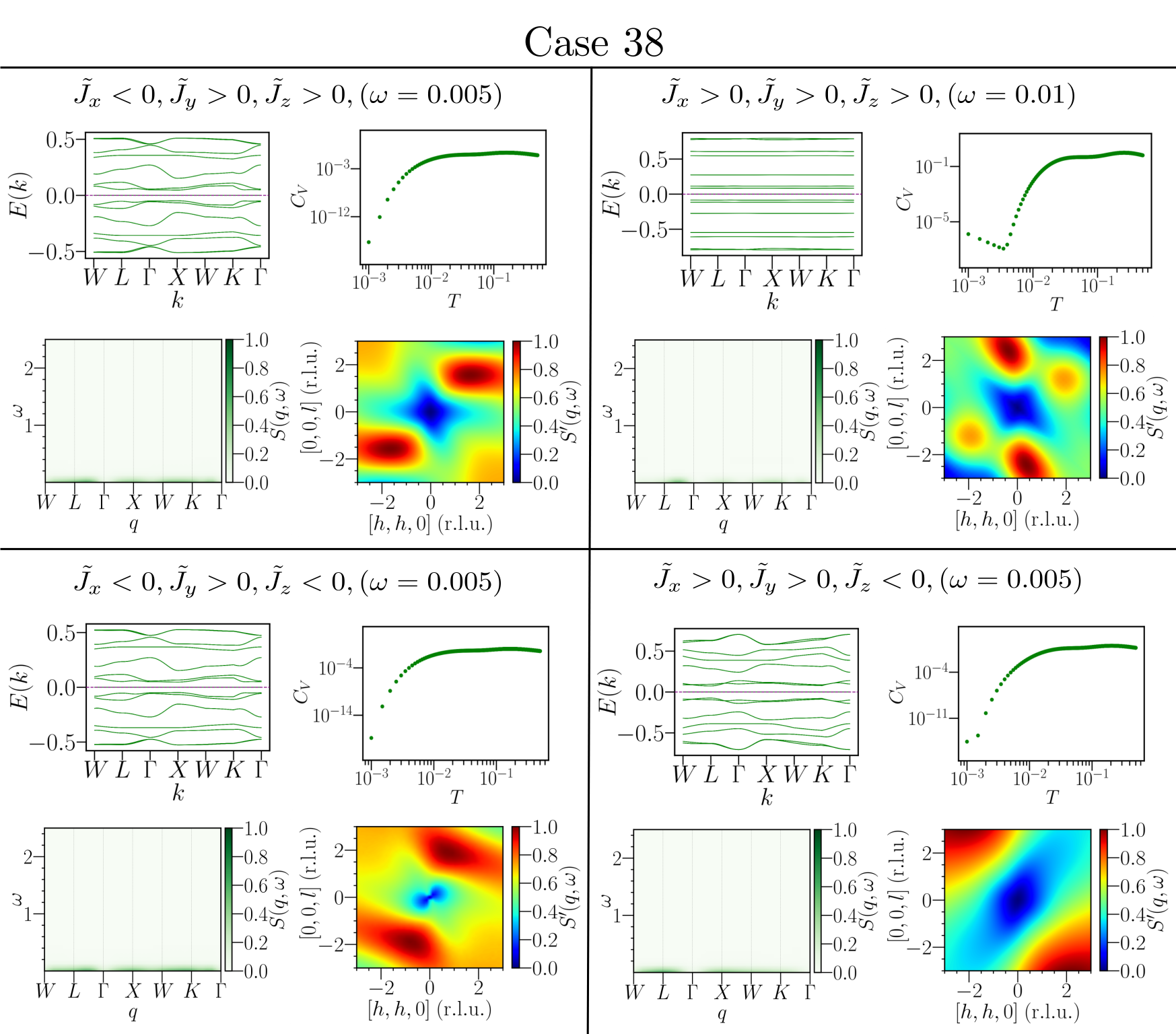}

    \caption{PSG case 38. A flat band at the Fermi energy coexists with other nearly flat, gapped bands in all four exchange coupling regimes. The low-energy $C_V(T)$ does not exhibit any power law in any temperature decade. The $S(\mathbf{q},\omega)$ weight is mostly confined to $\omega \sim 0.005 \tilde{J}$. The $S'(\mathbf{q},\omega)$ does not produce any special feature.}
    \label{case38}
\end{figure*}

The dynamical and static structure factors are evaluated across four distinct regions of the phase diagram that encompass all mean-field channels. An inspection of the phase diagram reveals that only four distinct cases realize the energy minimum under a systematic variation of the mean-field Hamiltonian's interaction parameters. Here, we investigate the dependence of the dispersion relation, low- temperature specific heat scaling and structure factor on the respective regions of the phase diagram for each of these four cases (see Fig.~\ref{case20} -~\ref{case38}). More specifically, four exchange-coupling regimes are considered: Regime I, $\tilde{J}_x>0$, $\tilde{J}_z>0$; Regime II, $\tilde{J}_x<0$, $\tilde{J}_z>0$; Regime III, $\tilde{J}_x<0$, $\tilde{J}_z<0$; and Regime IV, $\tilde{J}_x>0$, $\tilde{J}_z<0$, with $\tilde{J}_y=1$ fixed throughout. For each regime, the spinon band structure $E(\mathbf{k})$ and dynamical structure factor $S(\mathbf{q},\omega)$ are shown along the high-symmetry path $W$--$L$--$\Gamma$--$X$--$W$--$K$--$\Gamma$, together with the specific heat $C_V(T)$ and its low-temperature power-law fit $C_V\sim T^\gamma$, and the energy-resolved static structure factor $S'(\mathbf{q},\omega)$ in the $[h,h,l]$ plane at the indicated energy.

\section{Experimental Study}\label{sec:experiment}

We now discuss how the classified states connect to the two families of DO pyrochlore materials. The phase diagrams of Fig.~\ref{phase_diagram} are drawn in the $(\tilde{J}_x,\tilde{J}_z)$ plane at fixed $\tilde{J}_y$. Because the rotation to the XYZ form is taken about the local $y$ axis (Sec.~\ref{sec:hamiltonian}), the component $\tau^{y}=S^{y}$ is left invariant, so $\tilde{J}_y=J_{yy}$ is the octupolar coupling while $\tilde{J}_x$ and $\tilde{J}_z$ are the (rotated) dipolar couplings. Increasing $\tilde{J}_y$ relative to $\tilde{J}_x,\tilde{J}_z$ therefore tunes toward the octupolar-dominated regime. Within this plane the two DO families occupy distinct regions. The Ce-based compounds Ce$_2$Zr$_2$O$_7$, Ce$_2$Sn$_2$O$_7$, and Ce$_2$Hf$_2$O$_7$ remain disordered and are described as dipolar or octupolar $U(1)$ QSI proximate to the antiferromagnetic (all-positive) region~\cite{gaudet_quantum_2019,gao_experimental_2019,smith_case_2022,sibille_quantum_2020,bhardwaj_sleuthing_2022}, whereas the Nd-based compounds Nd$_2$Zr$_2$O$_7$ and Nd$_2$ScNbO$_7$ order into AIAO with dynamical fragmentation and lie in the region whose classical parent is that order~\cite{lhotel_fluctuations_2015,xu_anisotropic_2019,scheie_beyond_2021,benton_ground_2015,smith_experimental_2025}. A direct comparison with the bosonic mean-field phase diagram of Ref.~\cite{desrochers_competing_2022} places the fermionic $Z_2$ states obtained here alongside their bosonic counterparts; substituting the measured exchange parameters of the individual compounds into the phase diagram identifies the candidate PSG state for each material.

The low-temperature specific heat provides the sharpest fermionic diagnostic. For a gapless spinon spectrum in three dimensions, the leading power in $C_V$ is controlled by the geometry of the manifold on which the Bogoliubov energy vanishes: isolated (point) nodes give $C_V\sim T^3$, line nodes give $C_V\sim T^2$, and a finite density of states at the Fermi level---a spinon Fermi surface, or a band that is flat at the Fermi level---gives $C_V\sim T$ (and, when the density of states diverges, a still weaker power). The exponents in Table~\ref{MFH_summary} should therefore be read as fingerprints of the nodal structure of each ansatz rather than as isolated numbers. In particular, the near-linear behavior of case 19 ($\gamma\approx1.07$) is the fermionic-spinon analogue of the anomalous $T$-linear specific heat reported in Nd$_2$ScNbO$_7$~\cite{scheie_beyond_2021} and realizes the same mechanism identified in the fermionic $U(1)$ study of the DO model~\cite{sahu_u1_2024}; the sub-linear exponents ($\gamma<1$) are associated with bands flat at the Fermi level and are discussed further, together with their stability, in Sec.~\ref{sec:conclusion}. We caution that in Nd$_2$ScNbO$_7$ the anomalous term coexists with $B$-site disorder and random fields~\cite{scheie_beyond_2021}, so a definitive spinon interpretation requires separating intrinsic from disorder contributions, for instance through the field and wavevector dependence of the continuum.

Finally, the dynamical structure factor probes only the dipole moment of the magnetization. Because the octupolar component $S^{y}$ carries no magnetic dipole moment, its coupling to the neutron vanishes, $g_y=0$, and we retain only $g_x,g_z$ in the $g$-tensor (Appendix~\ref{sec:local_global}). Consequently the $E^{y}$ and $D^{y}$ correlations built from $\tau^{y}$ are invisible to inelastic neutron scattering, and any DO QSL whose dominant channel is octupolar (large $\tilde{J}_y$, i.e.\ $D^{y}/E^{y}$-dominated states such as those winning the octupolar regime of Fig.~\ref{phase_diagram}) is neutron-dark in its octupolar sector---the situation invoked for the octupolar liquid Ce$_2$Sn$_2$O$_7$~\cite{sibille_quantum_2020}. The dipolar-active states, by contrast, display the broad continua of Figs.~\ref{st_factor} and \ref{case20}--\ref{case38}; the continuum observed in Ce$_2$Zr$_2$O$_7$~\cite{gaudet_quantum_2019} is of the type expected from the gapless dipolar states identified here. Recent thermodynamic and neutron studies of DO pyrochlores~\cite{leger_impact_2024,luo_dynamical_2026,smith_quantum_2023} provide additional data against which these predictions can be tested.

\section{Conclusion}\label{sec:conclusion}

We have classified the symmetric $Z_2$ quantum spin liquids of fermionic partons on the dipolar-octupolar (DO) pyrochlore lattice and solved the corresponding mean-field theory across the anisotropic XYZ phase diagram. The central structural result is that the DO symmetry leaves the gauge structure of the classification unchanged---the $48$ symmetric $Z_2$ classes coincide with the generic spin--orbit-coupled result of Ref.~\cite{liu_symmetric_2021}, since the projective algebra is fixed by the space group and time reversal---but alters its physical realization: the DO symmetry action $U_{O}$, sublattice-independent with $U_S=-i\sigma^{y}$, selects a distinct set of symmetry-allowed hopping and pairing channels for each class, so that the DO fermionic $Z_2$ problem is a genuinely separate classification rather than a relabeling of the effective spin-$\tfrac{1}{2}$ one, with distinct spinon spectra and observables. Minimizing the mean-field energy over all symmetric singlet and triplet channels, we found that only a handful of $Z_2$ states are stabilized across the $(\tilde{J}_x,\tilde{J}_z)$ plane, and that these energy-minimizing states are gapless, carrying nodal or flat-band spinon spectra and---because each retains a nonzero triplet-pairing channel---a genuine $Z_2$ gauge structure. Together with the bosonic Schwinger-boson classification~\cite{desrochers_competing_2022}, the fermionic uniform-$U(1)$ analysis~\cite{sahu_u1_2024}, and the pseudofermion functional-renormalization-group phase diagram of the same model~\cite{chern_pseudofermion_2024}, the present work supplies the fermionic $Z_2$ entry in the parton description of DO pyrochlore magnets---the sector in which, unlike the $U(1)$ Coulomb regime, the topological order is realized equally naturally by fermionic and bosonic spinons.

Beyond the classification, these states carry distinct fingerprints, detailed in Sec.~\ref{sec:experiment}, that a gapped-photon quantum spin ice cannot reproduce: a specific-heat exponent set by the spinon nodal structure (with a near-linear-in-$T$ form echoing the anomalous response of Nd$_2$ScNbO$_7$), broad spinon continua in the dynamical structure factor of the type reported for Ce$_2$Zr$_2$O$_7$, and a neutron-dark octupolar channel relevant to Ce$_2$Sn$_2$O$_7$. These signatures place the disordered Ce-based compounds and the AIAO-ordered but fractionalized-looking Nd-based compounds on a common footing within the fermionic $Z_2$ landscape.

Several caveats qualify this picture and set the agenda for future work. First, a PSG class fixes the projective symmetry action but not the gauge group realized by the self-consistent solution: a nominally $Z_2$ class whose energy minimum retains only hopping channels ($\chi$ and/or $E$) realizes an enlarged $U(1)$ invariant gauge group and is properly read as a $U(1)$ state. In the present phase diagrams the energy-minimizing states all retain a triplet-pairing ($D$) channel and are therefore genuine $Z_2$ liquids, so the classification of the winners is unaffected; but the label attached to each state in Table~\ref{MFH_summary} should be understood as the realized, not the nominal, gauge group, and the ``nodal $Z_2$'' identification should not be read onto the pairing-free minima.

Second, several of the energy-minimizing states possess spinon bands that are flat and pinned to the Fermi level. A flat band there implies a macroscopically degenerate filling and a diverging density of states, so the mean-field solution is fine-tuned: the half-filling constraint fixes the chemical potential only up to the flat-band degeneracy, and the state is unstable to arbitrarily weak residual interactions and gauge fluctuations, which will generically lift the degeneracy toward additional pairing or symmetry breaking. Whether these flat bands are protected by the projective symmetries or are accidental features of the nearest-neighbor ansatz that disperse once further-neighbor amplitudes are admitted remains to be settled; the sub-linear specific-heat exponents noted in Sec.~\ref{sec:observables} are a direct consequence of this flat-band density of states and should be read with the same caution.

Third, the mean-field treatment carries the usual variational caveats. The channel decomposition is not unique; we fixed the redundant coefficients to $A^{a}_{j}=-\tfrac{1}{4}$ in the isotropic limit, a choice that keeps the energy bounded from below and retains all channels, but the phase boundaries can in principle depend on the decoupling scheme. Moreover, the competing states are closely spaced in energy---the minima lie within $\approx 0.04\,\tilde{J}$ of one another---so the fine structure of Fig.~\ref{phase_diagram} should be read as a map of near-degenerate competitors rather than as sharp phase boundaries; the robust content is the identity of the small set of low-energy states and their persistence over a wide range of couplings. Finally, in three dimensions the choice of fermionic over bosonic partons is a physical hypothesis about spinon statistics rather than a calculational convenience---but a falsifiable one, since the two lead to different low-energy thermodynamics and spectra---while the classification itself is parton-agnostic, enumerating the complete list of symmetry-allowed fermionic states independent of which is ultimately selected in a given material.

These considerations point to several concrete directions. Incorporating gauge fluctuations beyond mean field, most directly through Gutzwiller-projected wavefunctions studied by variational Monte Carlo, as recently carried out for chiral pyrochlore spin liquids~\cite{liu_chiral_2026}---would sharpen the energetic competition among the near-degenerate states, test the stability of the flat-band ansätze, and allow correlation-based diagnostics that separate gauge-dominated Coulombic states from matter-dominated ones to be applied to the DO $Z_2$ liquids. The chiral extensions of the present symmetric classification, and the transitions out of these liquids by spinon pairing and condensation into the observed AIAO and other competing orders, can both be organized within the same PSG framework~\cite{liu_competing_2019}, linking the fermionic states directly to the ordered phases of the DO materials. Finally, substituting the measured exchange parameters of individual Ce- and Nd-based compounds into the phase diagram, benchmarking against the PFRG results for the same XYZ model~\cite{chern_pseudofermion_2024}, and computing the response to an applied field would enable a direct, material-specific confrontation with experiment---with the near-linear specific heat and spinon continua of the Nd family, and the neutron-dark octupolar sector of the Ce family, as the most immediate targets.

\begin{acknowledgments}
R.~D. and S.~S. acknowledge fruitful discussions with Yasir Iqbal. S.~S. acknowledges support from the Science and Engineering Research Board (Department of Science and Technology) Govt. of India, under grant no. SRG/2020/001525, Anusandhan National Research Foundation (Department of Science and Technology) Govt. of India, under grant no. CRG/2023/007457 and internal grant from the Indian Institute of Science Education and Research, Tirupati. 
\end{acknowledgments}

\FloatBarrier

\appendix

\section{Local and global transformation of spin}\label{sec:local_global}

The global  cartesian coordinate axes are defined as 
\[
\hat{x}=(1,0,0), \qquad 
\hat{y}=(0,1,0), \qquad 
\hat{z}=(0,0,1).
\]
The local coordinate system at each sublattice $\mu=0,1,2,3$ is given in Table~\ref{local_coordinate}. 
The transformation between local spin operators 
$S_{\mathbf{r}_\mu}$ 
and global spin operators 
$\tilde{S}_{\mathbf{r}_\mu}$ 
is implemented via the rotation matrices $R_\mu$ defined below.

\begin{table}[ht]
\centering
\renewcommand{\arraystretch}{1.4}
\caption{Local basis vectors for each of the four sublattices $\mu=0,1,2,3$.}
\label{local_table}
\begin{tabular}{c|cccc}
\hline\hline
$\mu$ & 0 & 1 & 2 & 3 \\
\hline
$\hat{x}_\mu$ 
& $-\frac{1}{\sqrt{6}}(-2,1,1)$ 
& $-\frac{1}{\sqrt{6}}(1,2,-1)$ 
& $-\frac{1}{\sqrt{6}}(-1,1,2)$ 
& $\frac{1}{\sqrt{6}}(1,1,2)$ \\

$\hat{y}_\mu$ 
& $\frac{1}{\sqrt{2}}(0,-1,1)$ 
& $\frac{1}{\sqrt{2}}(1,0,1)$ 
& $\frac{1}{\sqrt{2}}(1,1,0)$ 
& $\frac{1}{\sqrt{2}}(1,-1,0)$ \\

$\hat{z}_\mu$ 
& $\frac{1}{\sqrt{3}}(1,1,1)$ 
& $\frac{1}{\sqrt{3}}(-1,1,1)$ 
& $\frac{1}{\sqrt{3}}(1,-1,1)$ 
& $\frac{1}{\sqrt{3}}(1,1,-1)$ \\
\hline\hline
\end{tabular} \label{local_coordinate}
\end{table}

The corresponding rotation matrices $R_\mu$ relating local and global frames are

\[
{S}^{a}_{\mathbf{r}_\mu}
=
 R_{\mu}^{ab}
\tilde{S}^{b}_{\mathbf{r}_\mu}.
\]

Explicitly,

\[
R_0 =
\begin{pmatrix}
\sqrt{\frac{2}{3}} & -\frac{1}{\sqrt{6}} & -\frac{1}{\sqrt{6}} \\
0 & -\frac{1}{\sqrt{2}} & \frac{1}{\sqrt{2}} \\
-\frac{1}{\sqrt{3}} & -\frac{1}{\sqrt{3}} & -\frac{1}{\sqrt{3}}
\end{pmatrix},
\]

\[
R_1 =
\begin{pmatrix}
-\frac{1}{\sqrt{6}} & -\sqrt{\frac{2}{3}} & \frac{1}{\sqrt{6}} \\
\frac{1}{\sqrt{2}} & 0 & \frac{1}{\sqrt{2}} \\
-\frac{1}{\sqrt{3}} & \frac{1}{\sqrt{3}} & \frac{1}{\sqrt{3}}
\end{pmatrix},
\]

\[
R_2 =
\begin{pmatrix}
\frac{1}{\sqrt{6}} & -\frac{1}{\sqrt{6}} & -\sqrt{\frac{2}{3}} \\
\frac{1}{\sqrt{2}} & \frac{1}{\sqrt{2}} & 0 \\
\frac{1}{\sqrt{3}} & -\frac{1}{\sqrt{3}} & \frac{1}{\sqrt{3}}
\end{pmatrix},
\]

\[
R_3 =
\begin{pmatrix}
\frac{1}{\sqrt{6}} & \frac{1}{\sqrt{6}} & \sqrt{\frac{2}{3}} \\
\frac{1}{\sqrt{2}} & -\frac{1}{\sqrt{2}} & 0 \\
\frac{1}{\sqrt{3}} & \frac{1}{\sqrt{3}} & -\frac{1}{\sqrt{3}}
\end{pmatrix}
\]

The Hamiltonian in the local frame in the presence of an external magnetic field $\mathbf{h}$ is

\[
\begin{aligned}
H_L
&=
\sum_{\langle \mathbf{r}_\mu,\mathbf{r}'_\nu \rangle}
\
J_{xx} S^x_{\mathbf{r}_\mu} S^x_{\mathbf{r}'_\nu}
+
J_{yy} S^y_{\mathbf{r}_\mu} S^y_{\mathbf{r}'_\nu}
+
J_{zz} S^z_{\mathbf{r}_\mu} S^z_{\mathbf{r}'_\nu} \\
&\quad
+
J_{xz}
\left(
S^x_{\mathbf{r}_\mu} S^z_{\mathbf{r}'_\nu}
+
S^z_{\mathbf{r}_\mu} S^x_{\mathbf{r}'_\nu}
\right)
\
-
\sum_{\mathbf{r}_\mu}
(\hat{z}_\mu \cdot \mathbf{h})
\left(
g_x S^x_{\mathbf{r}_\mu}
+
g_z S^z_{\mathbf{r}_\mu}
\right).
\end{aligned}
\]

In the global frame, the Hamiltonian takes the form

\[
H_G =
\sum_{\langle \mathbf{r}_\mu,\mathbf{r}'_\nu \rangle}
\sum_{a,b}
J^{ab}_{\mu\nu}
\tilde{S}^{a}_{\mathbf{r}_\mu}
\tilde{S}^{b}_{\mathbf{r}'_\nu}
-
\sum_{\mathbf{r}_\mu}
\sum_{a,b}
h^{a}
g^{ab}_{\mu}
\tilde{S}^{b}_{\mathbf{r}_\mu},
\]
where \( a,b \in \{x,y,z\} \) and \( \mu,\nu \in \{0,1,2,3\} \).

The exchange matrices \( J_{\mu\nu} \) are given by  \\

\begin{center}
    \begin{tabular}{c c}
        $
             J_{01}=\begin{pmatrix}
                 J_1 & J_2 & -J_1 \\
                 J_3 & -J_1 & J_4 \\
                 -J_4 & -J_1 & -J_3 \\
             \end{pmatrix}
        $
         & 
         $
             J_{02}=\begin{pmatrix}
                 -J_1 & J_1 & J_2 \\
                 J_4 & J_3 & -J_1 \\
                 -J_3 & -J_4 & -J_1 \\
             \end{pmatrix}
         $
         \\
        
         $
             J_{03}=\begin{pmatrix}
                 -J_1 & -J_1 & -J_2 \\
                 J_4 & -J_3 & J_1 \\
                 -J_3 & J_4 & J_1 \\
             \end{pmatrix}
         $
         & 
        $
             J_{12}=\begin{pmatrix}
                 -J_3 & -J_4 & -J_1 \\
                 J_1 & -J_1 & -J_2 \\
                 -J_3 & -J_4 & J_1 \\
             \end{pmatrix}
         $
         \\
         
        $
             J_{13}=\begin{pmatrix}
                 -J_3 & J_4 & J_1 \\
                 J_1 & J_1 & J_2 \\
                 -J_4 & J_3& -J_1 \\
             \end{pmatrix}
         $
         & 
        $
             J_{23}=\begin{pmatrix}
                 -J_4 & J_4 & -J_1 \\
                 -J_3 & J_4 & J_1 \\
                 J_1 & J_1 & J_2 \\
             \end{pmatrix},
        $
         \\
    \end{tabular}
\end{center}

The coupling constants are

\[
\begin{aligned}
J_1 &= \frac{1}{6}(-2J_{xx}+2J_{zz}-\sqrt{2}J_{xz}), \\
J_2 &= \frac{1}{3}(-2J_{xx}-J_{zz}+2\sqrt{2}J_{xz}), \\
J_3 &= \frac{1}{6}(J_{xx}-3J_{yy}+2J_{zz}+2\sqrt{2}J_{xz}), \\
J_4 &= \frac{1}{6}(-J_{xx}-3J_{yy}-2J_{zz}-2\sqrt{2}J_{xz}).
\end{aligned}
\]

 $g$-tensors are

\[
g_+ = \frac{1}{3}\left(\frac{g_x}{\sqrt{2}} + g_z\right),
\qquad
g_- = \frac{1}{3}\left(\sqrt{2} g_x - g_z\right).
\]

\[
g_0 =
\begin{pmatrix}
- g_- & g_+ & g_+ \\
- g_- & g_+ & g_+ \\
- g_- & g_+ & g_+
\end{pmatrix},
\quad
g_1 =
\begin{pmatrix}
g_+ & g_- & -g_+ \\
- g_+ & -g_- & g_+ \\
- g_+ & -g_- & g_+
\end{pmatrix},
\]

\[
g_2 =
\begin{pmatrix}
g_+ & -g_+ & -g_- \\
- g_+ & g_+ & g_- \\
g_+ & -g_+ & -g_-
\end{pmatrix},
\quad
g_3 =
\begin{pmatrix}
g_+ & g_+ & g_- \\
g_+ & g_+ & g_- \\
- g_+ & -g_+ & -g_-
\end{pmatrix}.
\]
 In our calculation we consider $g_{x}=-0.2324$ ,  $g_{z}=2.35$ and $g_{y}=0$.

\section{Dipolar-Octupolar symmetry}\label{subsec:do_symmetry}

The $D_{3d}$ site symmetry of the pyrochlore lattice has inversion, three $C_2$ rotations, a $C_3$ rotation along with various combinations of these operations. The ground-state manifold of a single ion is described by the quantum number $J$ of total angular momentum. The degeneracy of the ground-state manifold is lifted by the surrounding crystal electric field (CFT), and the lowest energy states form a doublet ($\ket{\pm}$). Note that $S^z=\frac{\ket{+}\bra{+}~-~\ket{-}\bra{-}}{2}$ and $S^{\pm}=\ket{\pm}\bra{\mp}$ are the definitions of the pseudo spin operators. The pseudo spin operators transform under the space group
symmetry operations. By mixing the sublattices, the symmetry operation makes a change in the local coordinate system. Thus, the pseudo-spin transformation has two unitary operations: one that accounts for the symmetry operation and another that monitors the change in the local coordinate system caused by sublattice mixing.

The $\bar{C}_{6}$ and $S$ operation on $\ket{\pm}$ is achieved with
\begin{equation*}
    R_{\bar{C}_{6},\mu} = e^{-i\frac{2\pi}{3}\hat{n}_{\bar{C}_{6},\mu}.\Vec{J}}, ~~ R_{S,\mu} = e^{-i\pi\hat{n}_{S,\mu}.\Vec{J}}.
\end{equation*}
Where, the $\hat{n}_{\mathcal{O},\mu}$ denotes the axes of rotation in local frame at $\mu$th sublattice. In global frame, $\hat{n}_{\bar{C}_{6}}=\frac{1}{\sqrt{3}}(1,1,1)$ and $\hat{n}_{S}=\frac{1}{\sqrt{2}}(1,1,0)$. 

The change in the local coordinate system is accounted by 
\begin{equation*}
    R_{\mu \rightarrow \nu} = e^{-i\theta_\alpha J_z} e^{-i\theta_\beta J_y} e^{-i\theta_\gamma J_z}.
\end{equation*}
Where $\theta_\alpha, \theta_\beta, \theta_\gamma$ are the Euler angles that takes from $\mu$th to $\nu$th local-coordinate system. 

The full transformation is given by
\begin{equation*}
    U_{\mathcal{O},\mu} = \mathcal{P}R_{\mu \rightarrow \mathcal{O}(\mu)} R_{\bar{C}_{6},\mu} \mathcal{P}
\end{equation*}
Where $\mathcal{P}$ is the projection operator which project into $\ket{\pm}$. The explicit calculation of the above expression leads to the following form of the $U_{\mathcal{O}}$ operators for DO spin:
\[
U_{T_i}=1_{2\times2},\quad
U_{\bar{C}_6}=1_{2\times2},\quad
U_S=-i\sigma_y,\quad
U_{\mathcal{T}}=i\sigma_y
\]
Note the $U_{\mathcal{O}}$ operators are all sublattice independent which is a characteristic feature of DO spins.

\section{ Mean-field decomposition} \label{MF decomposition}

We now demonstrate how the spin Hamiltonian can be cast in terms of various fermionic mean-field channels. Starting with the spin Hamiltonian (for notation simplicity we change the $\tilde{J}$ to $J$ and $\tau$ to $S$),
\begin{equation}\label{eq1}
    H = \sum_{\langle i,j \rangle}J_{ij}\mathbf{S}_i.\mathbf{S}_j
\end{equation}
Doing the parton construction,
\begin{equation}
\mathbf{S}_i=\frac{1}{2}f_{i\alpha}^{\dagger}\mathbf{\sigma}_{\alpha\beta}f_{i\beta}
\end{equation}
The spin Hamiltonian changes to a Hamiltonian with four-fermi terms. We want to write it in terms of all possible symmetry-preserving quadratic mean-field channels . We start by expressing the spin Hamiltonian,  in terms of the Fermionic operators in normal order :
\begin{equation}\label{eq28}
    \begin{split}
       & J_x S_i^x S_j^x = -\frac{J_x}{4}(f_{i\downarrow}^\dagger f_{j\downarrow}^\dagger f_{i\uparrow}f_{j\uparrow}+f_{i\downarrow}^\dagger f_{j\uparrow}^\dagger f_{i\uparrow}f_{j\downarrow}\\
       &+f_{i\uparrow}^\dagger f_{j\downarrow}^\dagger f_{i\downarrow}f_{j\uparrow}+f_{i\uparrow}^\dagger f_{j\uparrow}^\dagger f_{i\downarrow}f_{j\downarrow}) \\
       & J_y S_i^y S_j^y = \frac{J_y}{4}(f_{i\downarrow}^\dagger f_{j\downarrow}^\dagger f_{i\uparrow}f_{j\uparrow}-f_{i\downarrow}^\dagger f_{j\uparrow}^\dagger f_{i\uparrow}f_{j\downarrow}\\
       &-f_{i\uparrow}^\dagger f_{j\downarrow}^\dagger f_{i\downarrow}f_{j\uparrow}+f_{i\uparrow}^\dagger f_{j\uparrow}^\dagger f_{i\downarrow}f_{j\downarrow}) \\
       & J_z S_i^z S_j^z = \frac{J_z}{4}(-f_{i\uparrow}^\dagger f_{j\uparrow}^\dagger f_{i\uparrow}f_{j\uparrow}+f_{i\uparrow}^\dagger f_{j\downarrow}^\dagger f_{i\uparrow}f_{j\downarrow}\\
       &+f_{i\downarrow}^\dagger f_{j\uparrow}^\dagger f_{i\downarrow}f_{j\uparrow}-f_{i\downarrow}^\dagger f_{j\downarrow}^\dagger f_{i\downarrow}f_{j\downarrow}). \\
    \end{split}
\end{equation}
Consider the following generic expression where operators are the mean-field channels and write them in the same order in which the spin operators are written . 
\begin{equation}\label{eq29}
    \begin{split}
        & A_1\hat{\chi}_{ij}^\dagger\hat{\chi}_{ij}+A_2\hat{\Delta}_{ij}^\dagger\hat{\Delta}_{ij}+A_3{\hat{E}_{ij}}^{x\dagger }\hat{E}_{ij}^x+A_4{\hat{E}_{ij}}^{y\dagger} \hat{E}_{ij}^y\\
        &+A_5{\hat{E}_{ij}}^{z\dagger} \hat{E}_{ij}^z+A_6{\hat{D}_{ij}}^{x\dagger} \hat{D}_{ij}^x +A_7{\hat{D}_{ij}}^{y\dagger} \hat{D}_{ij}^y\\
        &+A_8{\hat{D}_{ij}}^{z\dagger} \hat{D}_{ij}^z+A_9\hat{n}_i\hat{n}_j+A_{10}\hat{n}_j \\
        & = \hat{n}_j(A_1+A_3+A_4+A_5+A_{10})\\
        & +(f_{i\uparrow}^\dagger f_{j\uparrow}^\dagger f_{i\uparrow}f_{j\uparrow}+f_{i\downarrow}^\dagger f_{j\downarrow}^\dagger f_{i\downarrow}f_{j\downarrow})(A_1+A_5-A_6-A_7-A_9)\\
        &+(f_{i\downarrow}^\dagger f_{j\uparrow}^\dagger f_{i\uparrow}f_{j\downarrow}+f_{i\uparrow}^\dagger f_{j\downarrow}^\dagger f_{i\downarrow}f_{j\uparrow})(A_1+A_2-A_5-A_8)\\
        & +(f_{i\downarrow}^\dagger f_{j\uparrow}^\dagger f_{i\downarrow}f_{j\uparrow}+f_{i\uparrow}^\dagger f_{j\downarrow}^\dagger f_{i\uparrow}f_{j\downarrow})(-A_2+A_3+A_4-A_8-A_9)\\
        & +(f_{i\uparrow}^\dagger f_{j\uparrow}^\dagger f_{i\downarrow}f_{j\downarrow}+f_{i\downarrow}^\dagger f_{j\downarrow}^\dagger f_{i\uparrow}f_{j\uparrow})(A_3-A_4+A_6-A_7)\\
    \end{split}
\end{equation}

\textbf{AFM coupling i.e. $\mathbf{J_x, J_y, J_z > 0}$}

In order to decompose the spin-spin interaction terms into the fermionic mean-field operator, the coefficients of the four-fermi terms in Eq.~\ref{eq28} are compared with the coefficients of four-fermi terms in Eq.~\ref{eq29}. 
\begin{equation}
\begin{split}
    & A_{10}^a=-(A_1^a+A_3^a+A_4^a+A_5^a)\\
    & A_9^a=-\frac{1}{4}-A_1^a-2A_2^a+A_3^a+A_4^a+A_5^a\\
    & A_8^a = \frac{1}{4}(1-\delta_{a,z})+A_1^a+A_2^a-A_5^a\\
    & A_7^a = \frac{1}{4}(1-\delta_{a,y})+A_1^a+A_2^a-A_4^a\\
    & A_6^a = \frac{1}{4}(1-\delta_{a,x})+A_1^a+A_2^a-A_3^a\\
\end{split}
\end{equation}
Where the superscript $a \in \{x,y,z\}$ to the coefficients A are put to distinguish the coefficients coming from $S_i^x S_j^x$, $S_i^y S_j^y$, $S_i^z S_j^z$.
Using above we can write
\begin{equation}\label{eq31}
    \begin{split}
    & ~~~~J_x S_i^x S_j^x+J_y S_i^y S_j^y +J_z S_i^z S_j^z \\
    & = \chi_{ij}^\dagger\chi_{ij}(J_x A_1^x +J_y A_1^y +J_z A_1^z)\\
    &+\Delta_{ij}^\dagger\Delta_{ij}(J_x A_2^x +J_y A_2^y +J_z A_2^z)\\
    & +{E_{ij}^x}^\dagger E_{ij}^x(J_x A_3^x +J_y A_3^y +J_z A_3^z)\\
    &+{E_{ij}^y}^\dagger E_{ij}^y(J_x A_4^x +J_y A_4^y +J_z A_4^z)\\
    &+{E_{ij}^z}^\dagger E_{ij}^z(J_x A_5^x +J_y A_5^y +J_z A_5^z) \\
    &+{D_{ij}^x}^\dagger D_{ij}^x[J_x(A_1^x+A_2^x-A_3^x)\\
    &+J_y(\frac{1}{4}+A_1^y+A_2^y-A_3^y)+J_z(\frac{1}{4}+A_1^z+A_2^z-A_3^z)]\\
    &+{D_{ij}^y}^\dagger D_{ij}^y[J_x(\frac{1}{4}+A_1^x+A_2^x-A_4^x)\\
    &+J_y(A_1^y+A_2^y-A_4^y)+J_z(\frac{1}{4}+A_1^z+A_2^z-A_4^z)]\\
    &+{D_{ij}^z}^\dagger D_{ij}^z[J_x(\frac{1}{4}+A_1^x+A_2^x-A_5^x)\\
    &+J_y(\frac{1}{4}+A_1^y+A_2^y-A_5^y)+J_z(A_1^z+A_2^z-A_5^z)]\\
    &+n_in_j[J_x(-\frac{1}{4}-A_1^x-2A_2^x+A_3^x+A_4^x+A_5^x)\\
    &+J_y(-\frac{1}{4}-A_1^y-2A_2^y+A_3^y+A_4^y+A_5^y)\\
    &+J_z(-\frac{1}{4}-A_1^z-2A_2^z+A_3^z+A_4^z+A_5^z)]\\
    &-n_j[J_x(A_1^x+A_3^x+A_4^x+A_5^x)\\
    &+J_y(A_1^y+A_3^y+A_4^y+A_5^y)\\
    &+J_z(A_1^z+A_3^z+A_4^z+A_5^z)] \\
    \end{split}
\end{equation}

\textbf{FM coupling i.e. $\mathbf{J_x, J_y, J_z < 0}$}

The spin Hamiltonian in terms of fermionic operators for the ferromagnetic case is
\begin{equation}\label{eq32}
\begin{aligned}
J_x S_i^x S_j^x &= \frac{|J_x|}{4} \Big(
f_{i\downarrow}^\dagger f_{j\downarrow}^\dagger f_{i\uparrow} f_{j\uparrow}
+ f_{i\downarrow}^\dagger f_{j\uparrow}^\dagger f_{i\uparrow} f_{j\downarrow} \\
&\quad + f_{i\uparrow}^\dagger f_{j\downarrow}^\dagger f_{i\downarrow} f_{j\uparrow}
+ f_{i\uparrow}^\dagger f_{j\uparrow}^\dagger f_{i\downarrow} f_{j\downarrow}
\Big),
\\[6pt]
J_y S_i^y S_j^y &= -\frac{|J_y|}{4} \Big(
f_{i\downarrow}^\dagger f_{j\downarrow}^\dagger f_{i\uparrow} f_{j\uparrow}
- f_{i\downarrow}^\dagger f_{j\uparrow}^\dagger f_{i\uparrow} f_{j\downarrow} \\
&\quad - f_{i\uparrow}^\dagger f_{j\downarrow}^\dagger f_{i\downarrow} f_{j\uparrow}
+ f_{i\uparrow}^\dagger f_{j\uparrow}^\dagger f_{i\downarrow} f_{j\downarrow}
\Big),
\\[6pt]
J_z S_i^z S_j^z &= -\frac{|J_z|}{4} \Big(
- f_{i\uparrow}^\dagger f_{j\uparrow}^\dagger f_{i\uparrow} f_{j\uparrow}
+ f_{i\uparrow}^\dagger f_{j\downarrow}^\dagger f_{i\uparrow} f_{j\downarrow} \\
&\quad + f_{i\downarrow}^\dagger f_{j\uparrow}^\dagger f_{i\downarrow} f_{j\uparrow}
- f_{i\downarrow}^\dagger f_{j\downarrow}^\dagger f_{i\downarrow} f_{j\downarrow}
\Big).
\end{aligned}
\end{equation}

Similarly comparing the coefficients of the four-fermi terms in Eq.~\ref{eq32} with the coefficients of four-fermi terms in Eq.~\ref{eq29} , we get

\begin{equation}
\begin{split}
    & A_{10}^a=-(A_1^a+A_3^a+A_4^a+A_5^a)\\
    & A_9^a=\frac{1}{4}-A_1^a-2A_2^a+A_3^a+A_4^a+A_5^a\\
    & A_8^a = -\frac{1}{4}(1-\delta_{a,z})+A_1^a+A_2^a-A_5^a\\
    & A_7^a = -\frac{1}{4}(1-\delta_{a,y})+A_1^a+A_2^a-A_4^a\\
    & A_6^a = -\frac{1}{4}(1-\delta_{a,x})+A_1^a+A_2^a-A_3^a\\
\end{split}
\end{equation}

Therefore, the most general mean-field decomposition for ferromagnetic coupling is

\begin{equation}\label{eq34}
    \begin{split}
    & ~~~~J_x S_i^x S_j^x+J_y S_i^y S_j^y +J_z S_i^z S_j^z \\
    & = \chi_{ij}^\dagger\chi_{ij}(J_x A_1^x +J_y A_1^y +J_z A_1^z)\\
    &+\Delta_{ij}^\dagger\Delta_{ij}(J_x A_2^x +J_y A_2^y +J_z A_2^z)\\
    & +{E_{ij}^x}^\dagger E_{ij}^x(J_x A_3^x +J_y A_3^y +J_z A_3^z)\\
    &+{E_{ij}^y}^\dagger E_{ij}^y(J_x A_4^x +J_y A_4^y +J_z A_4^z)\\
    &+{E_{ij}^z}^\dagger E_{ij}^z(J_x A_5^x +J_y A_5^y +J_z A_5^z) \\
    &+{D_{ij}^x}^\dagger D_{ij}^x[J_x(A_1^x+A_2^x-A_3^x)\\
    &+J_y(-\frac{1}{4}+A_1^y+A_2^y-A_3^y)+J_z(-\frac{1}{4}+A_1^z+A_2^z-A_3^z)]\\
    &+{D_{ij}^y}^\dagger D_{ij}^y[J_x(-\frac{1}{4}+A_1^x+A_2^x-A_4^x)\\
    &+J_y(A_1^y+A_2^y-A_4^y)+J_z(-\frac{1}{4}+A_1^z+A_2^z-A_4^z)]\\
    &+{D_{ij}^z}^\dagger D_{ij}^z[J_x(-\frac{1}{4}+A_1^x+A_2^x-A_5^x)\\
    &+J_y(-\frac{1}{4}+A_1^y+A_2^y-A_5^y)+J_z(A_1^z+A_2^z-A_5^z)]\\
    &+n_in_j[J_x(\frac{1}{4}-A_1^x-2A_2^x+A_3^x+A_4^x+A_5^x)\\
    &+J_y(\frac{1}{4}-A_1^y-2A_2^y+A_3^y+A_4^y+A_5^y)\\
    &+J_z(\frac{1}{4}-A_1^z-2A_2^z+A_3^z+A_4^z+A_5^z)]\\
    &-n_j[J_x(A_1^x+A_3^x+A_4^x+A_5^x)\\
    &+J_y(A_1^y+A_3^y+A_4^y+A_5^y)\\
    &+J_z(A_1^z+A_3^z+A_4^z+A_5^z)] \\
    \end{split}
\end{equation}

Eq.~\ref{eq31} and \ref{eq34} is the most general mean-field decomposition along different mean-field channels for AFM and FM cases.  \\

In the  isotropic limit ($J_x=J_y=J_z=J > 0$) , we have a wide array of  choices over the mean-field decomposition (given that the coefficients $A_j^i$, where $j ~\in~ \{1,2,3,4,5\}, i~\in~\{x,y,z\}$, can take any value). Here, we consider $A_j^i=-\frac{1}{4}$, which ensures the mean-field energy is bounded from below and the mean-field Hamiltonian consists of all possible mean-field channels. With this particular choice, the Hamiltonian is given by,
\begin{equation}
    \begin{split}
    H=&-\frac{3}{4}J\sum_{\langle i,j \rangle}[\{\hat{\chi}_{ij}^\dagger \hat{\chi}_{ij}+\hat{\Delta}_{ij}^\dagger \hat{\Delta}_{ij} +{\hat{E}_{ij}}^{x\dagger} \hat{E}_{ij}^x+{\hat{E}_{ij}}^{y\dagger} \hat{E}_{ij}^y\\
    &+{\hat{E}_{ij}}^{z\dagger} \hat{E}_{ij}^z+\frac{1}{3}({\hat{D}_{ij}}^{x\dagger} \hat{D}_{ij}^x+{\hat{D}_{ij}}^{y\dagger} \hat{D}_{ij}^y+{\hat{D}_{ij}}^{z\dagger} \hat{D}_{ij}^z)\}\\
    &-\frac{3}{4}\hat{n}_i\hat{n}_j+3\hat{n}_j]\\
    \end{split}
\end{equation}
Upon enforcing the half-filling, i.e.
\begin{equation*}
    f_{i\alpha}^\dagger f_{i\alpha}=1  
\end{equation*}                        

the last two terms will become constants and can be dropped, as this will shift the energy spectrum by a constant amount. Then we have , 
\begin{equation}
    \begin{split}
    H_{MF}&=-\frac{3}{4}J\sum_{\langle i,j \rangle}[\{{\chi}_{ij}^* \hat{\chi}_{ij}+{\Delta}_{ij}^* \hat{\Delta}_{ij} +{{E}_{ij}^x}^* \hat{E}_{ij}^x+{{E}_{ij}^y}^* \hat{E}_{ij}^y\\
    &+{{E}_{ij}^z}^* \hat{E}_{ij}^z+\frac{1}{3}({{D}_{ij}^x}^* \hat{D}_{ij}^x+{{D}_{ij}^y}^* \hat{D}_{ij}^y+{{D}_{ij}^z}^* \hat{D}_{ij}^z)+h.c.\}\\
    &-\{\abs{\chi_{ij}}^2+\abs{\Delta_{ij}}^2+\abs{E^x_{ij}}^2+\abs{E^y_{ij}}^2+\abs{E^z_{ij}}^2\\
    &+\frac{1}{3}(\abs{D^x_{ij}}^2+\abs{D^y_{ij}}^2+\abs{D^z_{ij}}^2)\}]\\
    \end{split}
\end{equation}    \\

\textbf{AFM coupling i.e. $\mathbf{J_x, J_y, J_z > 0}$}

In the first phase diagram, when only singlet and triplet hopping channels are present, we set the pairing term coefficients to zero, which imposes a constraint

\begin{align*}
& J_x A_2^x + J_y A_2^y + J_z A_2^z =0\\[4pt]
&J_x (A_1^x + A_2^x - A_3^x)
    + J_y \left(\frac{1}{4} + A_1^y + A_2^y - A_3^y\right) \\
  &\quad + J_z \left(\frac{1}{4} + A_1^z + A_2^z - A_3^z\right)=0 \\[4pt]
&J_x \left(\frac{1}{4} + A_1^x + A_2^x - A_4^x\right)
     + J_y (A_1^y + A_2^y - A_4^y)\\
  &\quad + J_z \left(\frac{1}{4} + A_1^z + A_2^z - A_4^z\right)=0 \\[4pt]
&J_x \left(\frac{1}{4} + A_1^x + A_2^x - A_5^x\right)
     + J_y \left(\frac{1}{4} + A_1^y + A_2^y - A_5^y\right) \\
  &\quad + J_z (A_1^z + A_2^z - A_5^z)=0
\end{align*}

Rearranging the above, we get the coefficients for all the triplet channels. As the MF Hamiltonian becomes anisotropic, we need to include the terms $ n_in_j $ and $n_j$, which also preserve the constraint relation. Consequently, the MF Hamiltonian takes the form,

\begin{equation}
\label{siglet_triplet}
\begin{split}
H &= \sum_{\langle i,j\rangle} \Bigg(
\chi_{ij}^\dagger\chi_{ij}
(J_xA_1^x+J_yA_1^y+J_zA_1^z)\\
&\quad+{E_{ij}^x}^\dagger E_{ij}^x
\Big[J_xA_1^x+J_yA_1^y+J_zA_1^z
+\frac14(J_y+J_z)\Big]\\
&\quad+{E_{ij}^y}^\dagger E_{ij}^y
\Big[J_xA_1^x+J_yA_1^y+J_zA_1^z
+\frac14(J_x+J_z)\Big]\\
&\quad+{E_{ij}^z}^\dagger E_{ij}^z
\Big[J_xA_1^x+J_yA_1^y+J_zA_1^z
+\frac14(J_x+J_y)\Big]\\
&\quad+h.c.\\
&\quad+n_in_j\Big[
-\frac14(J_x+J_y+J_z) - (J_xA_1^x+J_yA_1^y+J_zA_1^z)\\
&\qquad\quad
+(J_xA_1^x+J_yA_1^y+J_zA_1^z
+\frac14(J_y+J_z))\\
&\qquad\quad
+(J_xA_1^x+J_yA_1^y+J_zA_1^z
+\frac14(J_x+J_z))\\
&\qquad\quad
+(J_xA_1^x+J_yA_1^y+J_zA_1^z
+\frac14(J_x+J_y))
\Big]\\
&\quad-n_j\Big[
(J_xA_1^x+J_yA_1^y+J_zA_1^z)\\
&\qquad\quad
+(J_xA_1^x+J_yA_1^y+J_zA_1^z
+\frac14(J_y+J_z))\\
&\qquad\quad
+(J_xA_1^x+J_yA_1^y+J_zA_1^z
+\frac14(J_x+J_z))\\
&\qquad\quad
+(J_xA_1^x+J_yA_1^y+J_zA_1^z
+\frac14(J_x+J_y))
\Big]
\Bigg).
\end{split}
\end{equation}

 When only the singlet hopping and pairing channels are present, setting the triplet hopping and pairing coefficients to zero and solving recovers the isotropic constraint $J_x=J_y=J_z$. Within the singlet-triplet hopping phase diagram—where $J_x$ and $J_z$ vary from positive to negative values, keeping $J_y>0$ fixed—the Hamiltonian in Eq.~\ref{siglet_triplet} takes the form  \\

\textbf{$\mathbf{J_x>0, J_y>0, J_z>0, A_{1}^{x,y,z}= -\frac{1}{4}}$}
\begin{equation}
\begin{split}
H &= \sum_{\langle i,j \rangle}\Big[
\frac{1}{4}(-J_x-J_y-J_z)\,\chi_{ij}^{\dagger}\chi_{ij} \\
&\quad + \frac{1}{4}(-J_x)
E_{ij}^{x\dagger}E_{ij}^{x} 
+ \frac{1}{4}(-J_y) E_{ij}^{y\dagger}E_{ij}^{y} \\
&\quad + \frac{1}{4}(-J_z)
E_{ij}^{z\dagger}E_{ij}^{z}  + h.c \\
&\quad - \frac{1}{4}(-J_x-J_y-J_z) \Big].
\end{split}
\end{equation}              \\

\textbf{$\mathbf{J_x<0, J_y>0, J_z<0, A_{1}^{x,z}= \frac{1}{4} , A_{1}^{y}=-\frac{1}{4}}$}
\begin{equation}
\begin{split}
H &= \sum_{\langle i,j \rangle}\Big[
\frac{1}{4}(J_x-J_y+J_z)\,\chi_{ij}^{\dagger}\chi_{ij} \\
&\quad + \frac{1}{4}(J_x)
E_{ij}^{x\dagger}E_{ij}^{x}  + \frac{1}{4}(-J_y)
E_{ij}^{y\dagger}E_{ij}^{y} \\
&\quad + \frac{1}{4}(J_z)
E_{ij}^{z\dagger}E_{ij}^{z}  + h.c \\
&\quad - \frac{1}{4}(J_x-J_y+J_z) \Big] .
\end{split}
\end{equation}              \\

\textbf{$\mathbf{J_x<0, J_y>0, J_z>0, A_{1}^{x}= \frac{1}{4} ,A_{1}^{y}=-\frac{1}{4},A_{1}^{z}= -\frac{1}{4}} $}
\begin{equation}
\begin{split}
H &= \sum_{\langle i,j \rangle}\Big[
\frac{1}{4}(J_x-J_y-J_z)\,\chi_{ij}^{\dagger}\chi_{ij} \\
&\quad + \frac{1}{4}(J_x)
E_{ij}^{x\dagger}E_{ij}^{x} 
 + \frac{1}{4}(-J_y)
E_{ij}^{y\dagger}E_{ij}^{y} \\
&\quad + \frac{1}{4}(-J_z)
E_{ij}^{z\dagger}E_{ij}^{z}  + h.c \\
&\quad - \frac{1}{4}(J_x-J_y-J_z) \Big].
\end{split}
\end{equation}              \\

\textbf{$\mathbf{J_x>0, J_y>0, J_z<0, A_{1}^{x}= -\frac{1}{4} ,A_{1}^{y}=-\frac{1}{4},A_{1}^{z}= \frac{1}{4}}$}
\begin{equation}
\begin{split}
H &= \sum_{\langle i,j \rangle}\Big[
\frac{1}{4}(-J_x-J_y+J_z)\,\chi_{ij}^{\dagger}\chi_{ij} \\
&\quad + \frac{1}{4}(-J_x)
E_{ij}^{x\dagger}E_{ij}^{x} 
 + \frac{1}{4}(-J_y)
E_{ij}^{y\dagger}E_{ij}^{y} \\
&\quad + \frac{1}{4}(J_z)
E_{ij}^{z\dagger}E_{ij}^{z}  + h.c \\
&\quad - \frac{1}{4}(-J_x-J_y+J_z) \Big] .
\end{split}
\end{equation}              

We extended this analysis by evaluating the same phase diagram with all channels present. In this case, the Hamiltonian takes the form  \\

\textbf{$\mathbf{J_x>0, J_y>0, J_z>0, A_{j}^{x,y,z}= -\frac{1}{4}}$}
\begin{equation}
\begin{split}
H &= \sum_{\langle i,j \rangle}\Big[
\frac{1}{4}(-J_x-J_y-J_z)\,(\chi_{ij}^{\dagger}\chi_{ij}  + \Delta_{ij}^{\dagger}\Delta_{ij} + E_{ij}^{x\dagger}E_{ij}^{x} \\
& + E_{ij}^{y\dagger}E_{ij}^{y} + E_{ij}^{z\dagger}E_{ij}^{z}) + \frac{1}{4}(-J_x) D_{ij}^{x\dagger}D_{ij}^{x} + \frac{1}{4}(-J_y) D_{ij}^{y\dagger}D_{ij}^{y} \\
& + \frac{1}{4}(-J_z) D_{ij}^{z\dagger}D_{ij}^{z}  + h.c   - \frac{3}{4}(-J_x-J_y-J_z) \Big] .
\end{split}
\end{equation}              \\

\textbf{$\mathbf{J_x<0, J_y>0, J_z<0, A_{j}^{x,z}= \frac{1}{4} , A_{j}^{y}=-\frac{1}{4}}$}
\begin{equation}
\begin{split}
H &= \sum_{\langle i,j \rangle}\Big[
\frac{1}{4}(J_x-J_y+J_z)\,(\chi_{ij}^{\dagger}\chi_{ij}  + \Delta_{ij}^{\dagger}\Delta_{ij} + E_{ij}^{x\dagger}E_{ij}^{x} \\
& + E_{ij}^{y\dagger}E_{ij}^{y} + E_{ij}^{z\dagger}E_{ij}^{z}) + \frac{1}{4}(J_x) D_{ij}^{x\dagger}D_{ij}^{x} + \frac{1}{4}(-J_y) D_{ij}^{y\dagger}D_{ij}^{y} \\
& + \frac{1}{4}(J_z) D_{ij}^{z\dagger}D_{ij}^{z}  + h.c  - \frac{3}{4}(J_x-J_y+J_z) \Big] .
\end{split}
\end{equation}              \\

\textbf{$\mathbf{J_x<0, J_y>0, J_z>0, A_{j}^{x}= \frac{1}{4} ,A_{j}^{y}=-\frac{1}{4},A_{j}^{z}= -\frac{1}{4}} $}
\begin{equation}
\begin{split}
H &= \sum_{\langle i,j \rangle}\Big[
\frac{1}{4}(J_x-J_y-J_z)\,(\chi_{ij}^{\dagger}\chi_{ij}  + \Delta_{ij}^{\dagger}\Delta_{ij} + E_{ij}^{x\dagger}E_{ij}^{x} \\
& + E_{ij}^{y\dagger}E_{ij}^{y} + E_{ij}^{z\dagger}E_{ij}^{z}) + \frac{1}{4}(J_x) D_{ij}^{x\dagger}D_{ij}^{x} + \frac{1}{4}(-J_y) D_{ij}^{y\dagger}D_{ij}^{y} \\
& + \frac{1}{4}(-J_z) D_{ij}^{z\dagger}D_{ij}^{z}  + h.c   - \frac{3}{4}(J_x-J_y-J_z) \Big].
\end{split}
\end{equation}              \\

\textbf{$\mathbf{J_x>0, J_y>0, J_z<0, A_{j}^{x}= -\frac{1}{4} ,A_{j}^{y}=-\frac{1}{4},A_{j}^{z}= \frac{1}{4}}$}
\begin{equation}
\begin{split}
H &= \sum_{\langle i,j \rangle}\Big[
\frac{1}{4}(-J_x-J_y+J_z)\,(\chi_{ij}^{\dagger}\chi_{ij}  + \Delta_{ij}^{\dagger}\Delta_{ij} + E_{ij}^{x\dagger}E_{ij}^{x} \\
& + E_{ij}^{y\dagger}E_{ij}^{y} + E_{ij}^{z\dagger}E_{ij}^{z}) + \frac{1}{4}(-J_x) D_{ij}^{x\dagger}D_{ij}^{x} + \frac{1}{4}(-J_y) D_{ij}^{y\dagger}D_{ij}^{y} \\
& + \frac{1}{4}(J_z) D_{ij}^{z\dagger}D_{ij}^{z}  + h.c   - \frac{3}{4}(-J_x-J_y+J_z) \Big] .
\end{split}
\end{equation}              \\

\section{Symmetry relation between mean-field parameters}\label{Symmetry relation between mean-field parameters}

To determine the mean-field parameters for all bonds within the unit cell, we need to find the symmetry operation that maps a reference bond onto the remaining bonds inside the unit cell. The specific symmetry operations required for this are detailed below :

\[
\begin{aligned}
 (23) = \Sigma C_3 \Sigma C_3^{-1} \Sigma C_3^{-1}  :  (01) \rightarrow (01) 
\end{aligned}
\]    

\[
\begin{aligned}
 (14) = \Sigma C_3\Sigma C_3^{-1} \Sigma C_3^{-1} \Sigma C_3 \Sigma C_3  : (01) \rightarrow (10) 
\end{aligned}
\]   

\[
\begin{aligned}
 (14)(23) = \Sigma C_3\Sigma C_3  : (01) \rightarrow (10) 
\end{aligned}
\]   

\[
\begin{aligned}
 C_3 = \bar{C_6}^{-2} & : (01) \rightarrow (02) \\
    &  \quad \, (0\bar{1}) \rightarrow (0\bar{2}) \\
    &  \quad \, (12) \rightarrow (23) \\
    &  \quad \, (\bar{1}\bar{2}) \rightarrow (\bar{2} \bar{3}) 
\end{aligned}
\]

\[
\begin{aligned}
 C_3^{2} = \bar{C_6}^{2} & : (01) \rightarrow (03) \\
    &  \quad \, (0\bar{1}) \rightarrow (0\bar{3}) \\
    &  \quad \, (12) \rightarrow (31) \\
    &  \quad \, (\bar{1}\bar{2}) \rightarrow (\bar{3} \bar{1}) 
\end{aligned}
\]                           

\[
\begin{aligned}
 I = \bar{C_6}^{3} & : (01) \rightarrow (0 \bar{1}) \\
    &  \quad \, (12) \rightarrow (\bar{1} \bar{2}) \\
\end{aligned}
\]                           

\[
\begin{aligned}
 \Sigma = S.\mathcal{I} & : (01) \rightarrow (31) \\
\end{aligned}
\]

Where the right hand side depicts how the bonds are mapped within the unit cell. In our calculation, we used the bond $(01)$ in Fig.~\ref{pyrochlore_unit_cell} as our reference bond and mapped it to the other bonds within the pyrochlore unit cell. Under the $(23)$ operation the bond mapped to itself and under $(14)$ and $(14)(23)$ operation the bond getting flipped, i.e $(01) \rightarrow (10)$ . These operations will impose constraints on mean-field parameters, effectively reducing the mean-field parameter space.

\setcounter{figure}{8} 
\begin{figure}[h!]
    \centering
    \includegraphics[width=0.25\textwidth]{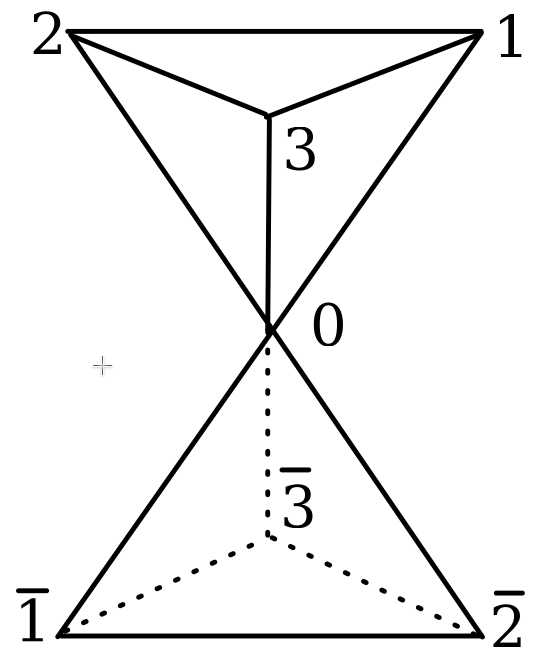}
    \caption{The pyrochlore unit cell with four sublattices, labeled as 0, 1, 2, and 3, and 12 bonds, is shown in the figure. The sublattices denoted with $\bar{1}$, $\bar{2}$, and $\bar{3}$, although they do not belong to the unit cell, are considered to label the bonds and to show the bond transformation upon symmetry operation.}
\label{pyrochlore_unit_cell}
\end{figure}

\setcounter{figure}{9} 
\begin{figure}[h!]
    \centering
    \includegraphics[width=0.45\textwidth]{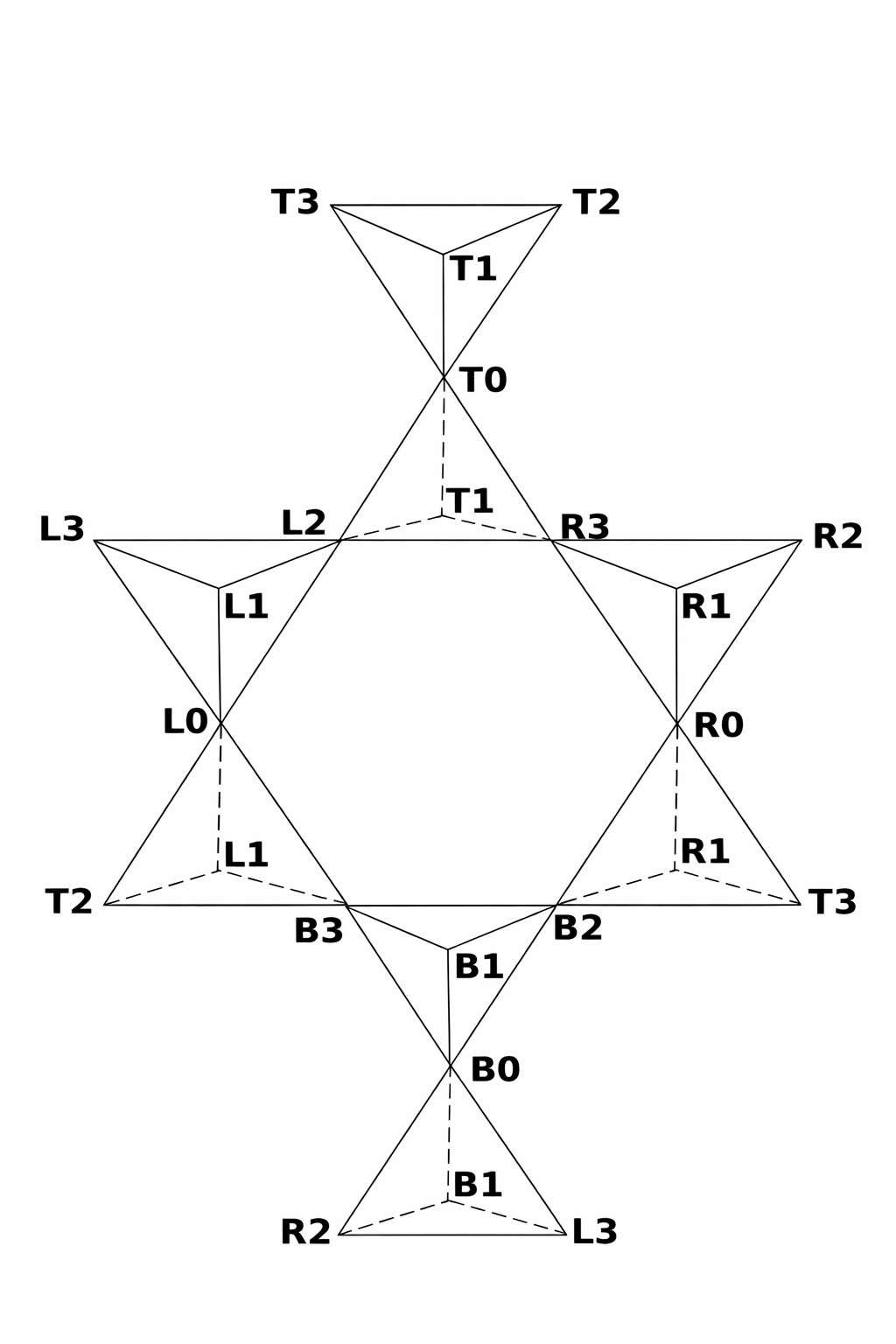}
    \caption{The extended pyrochlore unit cell has 16 sublattices. Each down-pointing tetrahedra consists of 4 sublattices, we denote them as bottom, right, top, and left.}
\label{extended_pyrochlore_unit_cell}
\end{figure}

We can use the translation PSG to establish the relation between different unit cells. When $\chi_{1}=0$, the mean fields are equal in every unit cell.

\[
u^\alpha_{T_i(i),\,T_i(j)} = u^\alpha_{i,j}, \quad i = 1,2,3.
\]

However, for $\chi_{1}=\pi$ ,

\[
T_1:\quad u^\alpha_{T_1(i),\,T_1(j)} = u^\alpha_{i,j},
\]
\[
T_2:\quad u^\alpha_{T_2(i),\,T_2(j)} = (-1)^{r_1 + r_1^{\prime}}\, u^\alpha_{i,j},
\]
\[
T_3:\quad u^\alpha_{T_3(i),\,T_3(j)} = (-1)^{r_1 + r_1^{\prime}} \, (-1)^{r_2 + r_2^{\prime}} \, u^\alpha_{i,j}.
\]

where $i=r_{\mu}=(r_1,r_2,r_3)_{\mu}$ and $j=r_{\nu}^{\prime}=(r_1^{\prime},r_2^{\prime},r_3^{\prime})_{\nu}$. Since the translation operations are defined as $ T_1 : \mathbf{r_\mu} = (r_1,r_2,r_3)_{\mu} \rightarrow (r_1+1,r_2,r_3)_\mu $, $ T_2 : \mathbf{r_\mu} = (r_1,r_2,r_3)_{\mu} \rightarrow (r_1,r_2+1,r_3)_\mu $ , $ T_3 : \mathbf{r_\mu}= (r_1,r_2,r_3)_{\mu} \rightarrow (r_1,r_2,r_3+1)_\mu $, doing a translation along $T_2$ and $T_3$ does not change $r_1$. Now, if we look at a unit cell for which $r_1=r_1^{\prime}$, the mean fields on these bonds do not pick up any negative sign upon doing a translation. However, for bonds with $r_1=r_1^{\prime} \pm 1$, upon a $T_2$ and $T_3$ translation, the mean field picks up a negative sign relative to the initial bond. Hence, we need to do two consecutive translations along $T_2$ and $T_3$ to restore the initial bond. Similarly, for bonds $r_2=r_2^{\prime} \pm 1$,  a $T_3$ translation changes the sign of the bond relative to the initial bond, but upon performing two consecutive $T_3$ operations, we again restore the initial bond. Therefore, for $\chi_{1}=\pi$, the unit cell became four time larger than the $\chi_{1}=0$ case. This is denoted as a larger unit cell (L) in Table~\ref{MFH_summary}. The extended pyrochlore unit cell consists of four down-pointing tetrahedra, labeled as bottom, right, top and left as shown in Fig.~\ref{extended_pyrochlore_unit_cell}. There are a total of 16 sublattices, four from each down-pointing tetrahedra, labeled $B0, B1, B2, B3; R0, R1, R2, R3; T0, T1, T2, T3;$ and $L0, L1, L2, L3$. 
\section{Energy minimization procedure}\label{Enery minimisation}
The mean-field Hamiltonian is defined as a function of the mean-field parameters (Eq.~\ref{meanfield-matrix}). The physical constraint of one particle per site is implemented as a Lagrange multiplier and we solve this mean-field Hamiltonian self-consistently to get the minimum energy. In a simple case, considering only singlet hopping and pairing, the zeroth-order mean-field Hamiltonian—subject to the one particle per site constraint—takes the form :

\begin{align}
H_{\mathrm{MF}} &=
-\frac{3J}{4} \sum_{\langle ij \rangle}
\Bigl[
\chi_{ij}\, f^{\dagger}_{i\alpha} f_{j\alpha}
+ \Delta_{ij}\, f^{\dagger}_{i\alpha} f^{\dagger}_{j\beta}
  \epsilon_{\alpha\beta}
+ \mathrm{H.c.}
\nonumber \\
&\hspace{3cm}
- |\chi_{ij}|^{2} - |\Delta_{ij}|^{2}
\Bigr]
\nonumber \\
&\quad
+ \sum_i
\
\lambda_0^3 \bigl( f^{\dagger}_{i\alpha} f_{i\alpha} - 1 \bigr)
+ \bigl(
(\lambda_0^1 + i \lambda_0^2)
f_{i\alpha} f_{i\beta}
\epsilon_{\alpha\beta}
+ \mathrm{H.c.}
\bigr)
\ 
\end{align}

Where $\chi_{ij}$, $\Delta_{ij}$ are the singlet hopping and pairing terms respectively and $\lambda$ is the Lagrange multiplier that fixes the chemical potential and coupling of particle and hole sectors. For the $Z_{2}$ spin liquid state on the pyrochlore lattice the Hamiltonian is represented in nambu (particle - hole)basis. The total mean-field energy is given by the sum of occupied eigenvalues over the Brillouin zone plus an extra quadratic term due to writing the fermion in a mean field basis (i.e, $|\chi_{ij}|^2$ and $|\Delta_{ij}|^2$). 

\begin{align}
E_{\text{total}}(\lambda_0^1,\lambda_0^2,\lambda_0^3, \chi_{ij}, \Delta_{ij})
&= \int_{\text{BZ}} \frac{d^3k}{(2\pi)^3}
   \sum_{\alpha,\, \omega_\alpha(\mathbf{k}) < 0} \omega_\alpha(\mathbf{k}) \nonumber \\
&\quad + \frac{9}{4} \big(|\chi_{ij}|^2 + |\Delta_{ij}|^2\big)
\end{align}

The integral is evaluated numerically using a midpoint Riemann sum method over a uniform grid in the Brillouin zone. For a fixed set of multipliers ($\lambda_{0},\lambda_{1}, \lambda_{2}$), the energy is minimized with respect to the variational parameter using the conjugate gradient method, and subsequently $
\frac{\partial E}{\partial \lambda_0^{1}}
= \frac{\partial E}{\partial \lambda_0^{2}}
= \frac{\partial E}{\partial \lambda_0^{3}}
= 0 $ were solved using the finite difference and root solver method. The minimization was repeated until the total energy converged to five decimal places to find the minimum energy.

\bibliographystyle{apsrev4-1}  
\bibliography{reference} 
\end{document}